\documentclass[a4paper, 11pt,reqno]{amsart} 
\usepackage{graphicx}
\usepackage{setspace}
\usepackage{dsfont}
\usepackage{amssymb, amsthm, amsmath, amstext} 
 \usepackage[utf8]{inputenc}
 \usepackage[T1]{fontenc} 
 \usepackage{cases}
 \usepackage{natbib} 

\usepackage{mathrsfs}
\newcommand{\ie}{\emph{i.e. }}

\newcommand{\ud}{\,\textup{d}}
 \newcommand{\mb}{\mathbf}
\newcommand{\point}{\,\cdot\,}
\newcommand{\PP}{\mathbb{P}}
\newcommand{\PE}{\mathbb{E}}
\newcommand{\PV}{\mathbb{V}\mathrm{ar}}
\newcommand{\dimens}{d}
\newcommand{\simpl}{\mb{ S}_{\dimens}}
\newcommand{\Ang}{\mathbf{w}} \newcommand{\Angrand}{\mathbf{W}}
\newcommand{\ang}{w} 
\newcommand{\Nu}{\boldsymbol{\nu}} %%{\mbox{\boldmath\(\nu\)}}
 
\newcommand{\wei}{\mathit{p}} \newcommand{\Wei}{\mathbf{p}}

\newcommand{\Mu}{\boldsymbol{\mu}}%{\mbox{\boldmath\(\mu\)}}
\newcommand{\pexc}{\zeta}
\newcommand{\Pexc}{\boldsymbol{\zeta}}%%{\mbox{\boldmath\(\zeta\)}}
\newcommand{\prob}{\mathbf{P}}
\newcommand{\thres}{v}
\newcommand{\bthres}{\mb{\thres}}
\newcommand{\fthres}{u}
\newcommand{\bfthres}{\boldsymbol{u}}%%{\mbox{\boldmath\(\upsilon\)}}
\newcommand{\margpar}{\chi}
\newcommand{\dmpar}{\psi}

\newcommand{\fullpar}{\theta}
\newcommand{\datatype}{\kappa}
\newcommand{\DM}{\textsc{DM}}

\newcommand{\MCMC}{\textsc{MCMC}}
\newcommand{\ndef}{n_{\text{det}}}
\newcommand{\nblock}{\mathcal{I}'}
\newcommand{\genemeas}{\ell}
\newcommand{\expmeas}{\lambda}
\newcommand{\leb}{\ell}
\newcommand{\hess}{\mathcal{H}}
\newcommand{\iter}{\iota}
\newcommand{\aug}{_+}
\DeclareMathOperator{\diri}{diri}
\DeclareMathOperator{\dbeta}{beta} 
\DeclareMathOperator{\PRM}{PP}

 \theoremstyle{plain} 
\newtheorem{proposition}{Proposition}
 \theoremstyle{remark} 
\newtheorem*{rem}{Remark}
\def\keywords{\vspace{.5em} {\textit{Keywords}:\, \relax%
  }}

\numberwithin{equation}{section}

\makeatletter
\newcommand{\authorfootnotes}{\renewcommand\thefootnote{\@fnsymbol\c@footnote}}%
 \makeatother
\usepackage{color}

\usepackage{url}

\begin{document}
\title[]{Semi-parametric  modeling  of 
excesses above  high  multivariate thresholds with  censored data}

\email{anne.sabourin@telecom-paristech.fr}%%sabourin@math.univ-lyon1.fr
\date{\today } 

%\maketitle
% \thispagestyle{empty}
% \pagestyle{empty}

\maketitle

\begin{center}

\authorfootnotes
  Anne Sabourin %\footnote{A.Sabourin}
\textsuperscript{1}
\par\bigskip
\textsuperscript{1} Institut Mines-Télécom, Télécom ParisTech, CNRS LTCI \\
%Département TSI,\\ 
37-38, rue Dareau, 75014 Paris, FRANCE\\
\texttt{anne.sabourin@telecom-paristech.fr}

\end{center}
\thispagestyle{empty}
\pagestyle{empty}

\begin{abstract}

  How to include censored data in a statistical analysis is a
recurrent issue in statistics. In multivariate extremes, the
dependence structure of large observations can be characterized in
terms of a non parametric \emph{angular measure}, while marginal
excesses above asymptotically large thresholds have a parametric
distribution.  In this work, a flexible semi-parametric Dirichlet
mixture model for angular measures is adapted to the context of
censored data and missing components. One major issue is to take into
account censoring intervals overlapping the extremal threshold,
without knowing whether the corresponding hidden data is actually
extreme. Further, the censored likelihood needed for Bayesian
inference has no analytic expression.  The first issue is tackled
using a Poisson process model for extremes, whereas a data
augmentation scheme avoids multivariate integration of the Poisson
process intensity over both the censored intervals and the failure
region above threshold. The implemented MCMC algorithm allows
simultaneous estimation of marginal and dependence parameters, so that
all sources of uncertainty other than model bias are captured by
posterior credible intervals. The method is illustrated on simulated
and real data.

\end{abstract}

%%\setpagewiselinenumbers
%%\modulolinenumbers[5]
%%\linenumbers
\section{Introduction}
\label{intro}

Data censoring is a commonly encountered problem in multivariate statistical analysis of
extreme values.   A
  `censored likelihood' approach 
makes it possible to take into account partially extreme data (non concomitant extremes): coordinates
that do not exceed some  large fixed threshold are simply considered
as left-censored. Thus, the possibly misleading information carried 
by non-extreme coordinates is ignored, only the fact that they are not
extreme is considered 
(\cite{smith1994multivariate, ledford1996statistics, smith1997markov},
see also \cite{thibaud2013efficient} or \cite{huser2014comparative}).
%First, observations may not be concomitantly extreme, 
 %  so that the marginal Pareto model does not apply to all the margins
 % forming a jointly extreme observation. 
% Multivariate analysis of extreme events %for environmental applications
% has received increased attention in the past
% decade. 
% I
However, there are other situations where the original data is
incomplete. % , which results in 
% left- or right-censoring.
For example,
%  in many applications,
% the scarcity of  extreme data  induces a large
% amount of uncertainty in the estimation of the marginal parameters
% and the dependence structure of extremes.
 one popular way to obtain large sample sizes in environmental
 sciences in general and in hydrology in particular, is to  take into
 account data reconstructed 
from archives, which
results in a certain amount of  left- and right-censored data, and
missing data. 
 As an example,  what originally  motivated this work is a
 hydrological data set consisting of
 % data set consists of
 daily  
 water discharge  recorded  at four neighboring stations % are
                                % available. %  
in
the region of the Gardons, in the south of France.
% Daily maxima have only recently been systematically recorded % (starting
 % from the $60's$)
 The extent of systematic recent records is short (a few decades)
 and varies from one
 station to another, so that standard inference using only `clean' data
 is unfeasible (% for reasonable extremal thresholds based on univariate
 % analysis, 
 only $3$ uncensored multivariate excesses of `large' thresholds
 - fixed after preliminary uni-variate analysis- are recorded).  
Historical information is available, % : unusually high water levels
% %occurring % before before the 1891 (the earliest one is dated from 1604)
% appear in
% the archives
starting from the $17^\text{th}$ century, a large part of it being
censored: only major floods are recorded, sometimes as an interval
data (\emph{e.g.} `the water level exceeded the parapet but the
Mr. X's house was spared'). These events are followed by long `blank'
periods during which the previous record was not
exceeded. % water level did not exceed
 % that of the previous flood.
 Uni-variate analysis for this data set has been carried on
 by \cite{neppel2010flood} but a multivariate analysis of extremes has
 never been accomplished, largely due  to the complexity of the data
 set, 
 with multiple censoring. % The aim of the present paper is to
 % propose  a new methodology for 
 % inferring the multivariate  structure of extremes using such censored data
 % sets. 

While  modeling multivariate extremes is a relatively
well marked out path when `exact' (non censored) data are at stake,
many fewer options are currently available for the statistician working
 with censored data. 
The aim of the present paper is to provide a flexible framework
allowing multivariate inference in this context.
 % Multivariate analysis of similar data, in a flexible 
 % dependence model, is the main aim of this paper. 
 Here, 
the focus is on the methodology and  the inferential framework is 
mainly  tested on simulated data
with a censoring pattern that resembles that of the real data. A 
 detailed 
 analysis of the hydrological data raises other issues, such as,
 among others, temporal dependence and added value of the most ancient data. These questions are addressed in a
 separate paper, intended for   the hydrological community
 \citep{sabourin:hal-01087687}\footnote{preprint available on \url{https://hal.archives-ouvertes.fr/hal-01087687}}.%  is reserved for 
 % a separate  study.

  % \citep[see \emph{e.g.}][and the references therein]{neppel2010flood}.

  Under a  standard assumption
%the assumption
 of multivariate regular variation (see Section~\ref{sec:model}),
% and  if the extremes are asymptotically dependent
% (that is, if unusually large realizations  tend to occur
%  simultaneously), the point process of suitably re-scaled
%  $\mathbb{R}^d$-valued observations on regions bounded
%  away from $\mb{0}_{\mathbb{R}^d}$ (\emph{i.e.}  of excesses above a multivariate high
%  threshold) is approximately Poisson \citep{resnick1987extreme,Resnick07}. % Its intensity can be
 % written in pseudo-polar coordinates $(r,\Ang)\;(r>0,\sum_{i=1}^d
 % w_j=1,w_j\ge 0)$ as   a product measure
 % of the form $c\frac{\ud r}{r^2}{\ud H(\Ang)}$.
 the distribution of excesses above large thresholds is characterized
 by parametric marginal distributions and a non-parametric dependence structure that
 is independent from threshold. 
Since the family of admissible dependence structures 
 is, by nature, too large  to be fully
 described by any parametric model,  non-parametric 
 estimation has  received a great deal of attention in the
 past few years % is  an important issue  
 % This  pleads in favour of non parametric inference
\citep{einmahl2001nonparametric,einmahl2009maximum,guillotte2011non}. % One
% drawback for uncertainty assessment is the absence of explicit expression for the variance of the
% maximum likelihood estimators. Further, t
To the best of my knowledge, the non parametric estimators of the
so-called \emph{angular measure} (which characterizes the dependence
structure among extremes) 
% spectral measure
are only defined with exact data and their adaptation
 to censored data is far from straightforward.

% hresholds  are  the dependence structure of extremes is entirely determined by an 
%  \emph{angular measure} $H$, which  is a finite measure on the
% unit sphere, and  only has to satisfy a first moments
% constraint. To wit, $H$ is the distribution of  the directional component of
% re-scaled observations above high thresholds. 
  For applied purposes, it is common practice to use a parametric
  dependence model. A  widely used one  is  the Logistic model and its
asymmetric and nested extensions \citep{gumbel1960distributions,
  coles1991modeling, stephenson2009high,
  stephenson2003simulating,fougeres2009models}. 
In the logistic family, % the dependence is expressed 
% \emph{via} the exponent function, which is an integral form of the
% angular measure.
%  The main advantage is that the
 censored
versions of the likelihood are readily available, but parameters are
subject to non linear constraints  and structural modeling choices
have to be made \emph{a priori}, \emph{e.g.}, by 
 allowing   only  bi-variate or tri-variate dependence between closest
 neighbors. 

One   semi-parametric compromise  consists in using  mixture models,
built from  a 
potentially infinite number of 
parametric components,  such as the  Dirichlet mixture model
(\DM), first  introduced by \cite{Boldi_Davison07}. 
They have shown  that it can approach
   arbitrarily well any valid angular measure for extremes. 
A re-parametrized
version of the $\DM$ model
 \citep{sabourinNaveau2012},   allows for  consistent Bayesian inference
 - thus,
 a straightforward uncertainty assessment using posterior credible
 sets - with a
 varying   number of mixture components \emph{via} a 
\emph{reversible-jumps}
 algorithm. The approach is appropriate  for  data sets of moderate
 dimension (typically, $d \approx 5$).

The  purpose of the present work is to adapt the \DM\   model 
 to the case of %multivariate 
censored data. %  and thus to provide a flexible
% inferential framework for this kind of data
 The  difficulties are 
two-fold: First, from a modeling perspective, when the censoring intervals overlap the extremal
thresholds (determined by preliminary  analysis), one cannot
tell whether the event must be treated as  extreme.  The proposed
approach here consists in  reformulating  the  
\emph{Peaks-over-threshold} (POT)  model originally proposed by
\cite{Boldi_Davison07} and \cite{sabourinNaveau2012}, 
  in terms of a \emph{Poisson
 model}, in which the censored regions overlapping the threshold have a
 well-defined likelihood. %  Indeed, the DM model was originally
 % formulated is \cite{Boldi_Davison07} and \cite{sabourinNaveau2012} as
 % a family of  generalized multivariate  Pareto distributions. 
 The second challenge is numerical and algorithmic:  for
right-censored  data above the extremal threshold (not overlapping it), 
the likelihood 
expression  involves  integrals of a density
 over rectangular regions, which have no analytic expression.  % Those
 % terms correspond  either to a missing data component or to a normalizing
 % constant involving the exponent measure of a failure region. 
 The
 latter  issue is  tackled within a  data augmentation framework,
 which is  implemented as an
 extension of \cite{sabourinNaveau2012}'s  algorithm
 for Dirichlet mixtures. % a rather classical 

An additional issue addressed in this paper concerns the separation
between  marginal parameters estimation and estimation of the
dependence structure.
 Performing the two steps separately is a widely used approach, but it boils   down to
neglecting marginal uncertainty, which confuses uncertainty assessment
about joint events such as probabilities of failure regions. It also
goes against the principle of  using  regional information together
with the dependence structure
 to improve marginal estimation, which is the main idea  of the popular  \emph{regional frequency analysis} in
 hydrology. 
In this paper, simultaneous inference of
marginal and dependence parameters in the \DM\ model is performed,
which amounts in practice to specifying additional steps for the
marginal parameters  in the MCMC sampler.

The rest of this paper is organized as follows: 
Section \ref{sec:model} recalls the necessary % probabilistic and
% statistical 
background
for  extreme values modeling. % A Poisson model for excesses is
% introduced, t
The   main features of the Dirichlet mixture model %  and its
% re-parametrised version
are
sketched. This  POT  model %   originally proposed in
% \cite{Boldi_Davison07} and \cite{sabourinNaveau2012}
is then reformulated as a Poisson
model, which addresses the issue of variable threshold induced by the
fluctuating marginal parameters. Censoring is  introduced in
Section~\ref{sec:censoredModel}. 
In this context,  the Poisson model has the additional advantage
that censored data overlapping threshold have a well defined
likelihood.%  However, % is also adapted to
% % censored data, which are %  %  the Poisson joint likelihood for regular,
% % % %  non censored data is
% % % % written.
% % % Censoring is
% censoring leads
% to a censored likelihood without analytic expression.
The lack of analytic expression for the latter is 
addressed  by 
 a  data augmentation scheme described in Section% the general 
% principles of data augmentation are recalled and a data augmentation
% scheme is proposed in Section
~\ref{sec:data_aug}.   % for the problem under
% study.
% Section~\ref{sec:MCMC
%   algo} describes the implementation of a MCMC algorithm which extends the   reversible-jump 
%  algorithm initially proposed by \cite{sabourinNaveau2012} 
% while incorporating two novel components (data augmentation and
% varying marginal parameters).
% implementing the above methods  in moderate dimension
% is described, which extends  previous work from
%  \cite{sabourinNaveau2012}
%  while allowing (with additional Gibbs steps)
%   inference of marginal parameters and data augmentation.
The method is illustrated by a simulation study in
Section~\ref{sec:results}: marginal performance in the \DM\ model and
in an independent one (without dependence structure) are compared, and
the predictive performance of the joint
model % scores in the dependence model
in terms of %with regards to
conditional probabilities of joint excesses is investigated. The model
is also fitted  to the hydrological data. Section
\ref{conclusion} concludes. Most of the technicalities needed for
practical implementation, such as computation of conditional
distributions, or details concerning the data augmentation scheme and
its consistency  are relegated to the appendix.

\section{Model for threshold excesses}\label{sec:model}
\subsection{Dependence structure model: angular measures}
In this paper, the sample space is the $d$-dimensional Euclidean space
$\mathbb R^d$, endowed with the Borel $\sigma$-field. In what
follows, bold symbols denote vectors and, unless otherwise mentioned,
binary operators applied to vectors are defined component-wise.
 Let $(\mb Y_t)_{t\in\mathbb N}$ be independent, 
identically
distributed (\emph{i.i.d.}) random vectors in $\mathbb R^d$, with
joint distribution $\mb F$ and margins $F_j$, $1\le j\le d$. The joint
behavior of large observations is best expressed in terms of
standardized data. 
 Namely,  define 

\noindent
$$% \label{eq:frechetTransformed}
\mb X_t = (-1/\log(F_1(Y_{1,t})), \dotsc, -1/\log(F_d(Y_{d,t})))  \,.
$$ %\end{equation}

\noindent 
Then %, if $\mb X$ is distributed as the $\mb X_t$'s,  
the $X_{j,t}$'s  have unit-Fréchet distribution, $\PP(X_{j,t}\le x) =
e^{-1/x}$, $x>0$. It is mathematically convenient to switch to pseudo-polar
coordinates, 

\noindent
$$
R = \sum_{j=1}^d X_{j}\;(\text{\emph{radial component}}),\quad \Angrand
= \frac{1}{R}\mb X \in \simpl \; (\text{\emph{angular component}}) \,,
$$

\noindent
where $\simpl = \{\mb x : x_j\ge 0, \sum_{j=1}^d x_j = 1\}$ is the
unit simplex. The radial
variable $R$ corresponds to the `amplitude' of the  data whereas the
angular component $\Angrand$ characterizes their `direction'.
Asymptotic theory
\citep{resnick1987extreme,Beirlant04,coles2001introduction} tells us
that, under mild assumptions on $\mb F$ (namely, belonging to a
multivariate maximum domain of attraction),
an appropriate  model, commonly referred to as a multivariate 
\emph{Peaks-over-threshold} (POT) model, for  $(R,\Angrand)$ over high radial
thresholds $r_0$,   is 

%\noindent
\begin{equation}
  \label{eq:angularModel}
\PP(R >r, \Angrand \in A \; | \; R>r_0 ) = \frac{r}{r_0}H(A)\;,
\qquad r_0>r, A\subset\simpl\,,  
\end{equation}

\noindent
where  $H$ is the so-called `angular probability measure' (called
 the `angular measure' in the remainder of this paper). The angular
measure is thus the limiting distribution of the angle, given that the
radius is large. % The constant  $c$ is a  normalising factor 
 % such that $H$ be a probability measure on $\simpl$.
Concentration of $H$'s  mass in
the middle of the simplex  indicates  strong
dependence at extreme levels, whereas Dirac masses only on the vertices
characterizes asymptotic independence.  This paper focuses  on the case where $H$
is concentrated on  the interior of the simplex, so that all
the variables are asymptotically dependent.

Because of the standardization to unit Fréchet, % standardization and to our choice of the
% $L_1$ norm, 
a probability measure $H$ on $\simpl$ is a valid angular
measure if and only if
$ 
% \noindent
% \begin{equation}\label{momentsBase}
\int_{\simpl} \ang_j \,\ud H(\Ang) = \frac{1}{d}\quad (1\le j\le
d)\,. 
% \end{equation}
% \noindent
$ This moments constraint  is the only condition on $H$, so that the
angular measure has no reason to be part of any particular parametric
family.  
\subsection{Dirichlet mixture angular measures}
In this paper, the angular measure $H$ is modeled by a  Dirichlet
mixture distribution \citep{Boldi_Davison07, sabourinNaveau2012}.
%  Angular measure densities will be
% given with respect to the $d-1$ dimensional Lebesgue measure
% $\ud\Ang=\ud\ang_1\dotsb\ud\ang_{d-1}$.
A Dirichlet distribution can be characterized by a  shape  $\nu\in
\mathbb{R}^+$ and a 
center of mass $\Mu \in \simpl$, so that its density 
% In the following, angular measure densities will be
% given
 with respect to the $d-1$ dimensional Lebesgue measure
$\ud\Ang=\ud\ang_1\dotsb\ud\ang_{d-1}$, is

\noindent
\begin{equation}\label{diriDensity}
\diri_{\nu,\Mu}(\Ang) = \frac{\Gamma(\nu)}{\prod_{j=1}^d
  \Gamma(\nu\mu_j)}
\prod_{j=1}^d \ang_j^{\nu\mu_j-1}\,\qquad (\Ang\in\simpl).
\end{equation}

\noindent
A parameter for a $k$-mixture is of the form

\noindent
 $$
\dmpar = \left( (\wei_1,\dotsc,\wei_k), (\Mu_{1},\dotsc,
   \Mu_{k} ), (\nu_1,\dotsc,\nu_k)
\right)\,,
$$
%%test $\boldsymbol{\mu}$ , $\Mu$, \mbox{\boldmath{${\mu}$}}

\noindent
 with weights $\wei_m>0$, such that  $\sum_{m=1}^k \wei_m =1$. This is summarized by
 writing   
 $\dmpar = \left( \wei_{1:k}, \Mu_{1:k}, \nu_{1:k}
\right)$.
 The corresponding
mixture density is 

\noindent 
\begin{equation}\label{dirimixDens}
h_{\dmpar}(\Ang) = \sum_{m=1}^k \wei_m
\diri_{\nu,\Mu_{m}}(\Ang)\;.
\end{equation}

\noindent
the moments constraint is satisfied if and only if 
 
\noindent
 \begin{equation*}%\label{moments}
\sum_{m=1}^k \wei_m \Mu_{m} = \left(1/\dimens,\dotsc,1/\dimens\right)\, ,
 \end{equation*}

\noindent
 which, in geometric terms, means that the center of mass of the
$\Mu_{1:k}$'s, with weights $\wei_{1:k}$, must lie at the center of
the simplex. As  established by 
\cite{Boldi_Davison07} and mentioned in the introduction, the family of Dirichlet
mixture densities satisfying  the moments constraint % \eqref{centerMassCondition} 
is weakly
dense in the space of admissible angular measure. 
In addition, in a Bayesian framework,
\cite{sabourinNaveau2012} have shown that the posterior is
weakly consistent under mild conditions.  These two features put 
together make the Dirichlet mixture model an adequate  candidate
for modeling the angular components of extremes. 
 
\subsection{Model for margins}\label{sec:marginalModel}
% Considering \eqref{cvPoisson}, 
The above  model for excesses  concerns standardized versions $\mb X_t$ of the
data $\mb Y_t$ involving  marginal cumulative distribution function $F_j$
($1\le j\le d$), which have to be estimated. 
As a  consequence of uni-variate extreme value theory %(Pickands' theorem,
\citep{pickands1975statistical},  uni-variate excesses above
large thresholds $v_j$ ($1\le j\le d$) are approximately distributed according to a
Generalized Pareto distribution with parameters $\xi_j$ (shape) and 
$\sigma_j$ (scale parameter), 

\noindent
$$
P(Y_j > y  \;|\; Y_j>v_j) \approx_{v_j\to \infty} (1+\xi_j \frac{y-v_j}{\sigma_j})^{-1/\xi_j}\,.
$$

\noindent
A widely used method to model the largest excesses  is a follows:  
Define a high multivariate threshold
$\bthres=(\thres_1,\dotsc,\thres_d)$ and
call `marginal
excess' any $Y_{j,t}>\thres_j$. Then, marginal excesses above $v_j$ are modeled
as
generalized Pareto random variables with parameters $\xi_j$ and
$\sigma_j$.  The marginal parameters are
gathered into a $(2 \dimens)$-dimensional vector

\noindent
$$
\margpar = \left(\log(\sigma_1),\dotsc,\log(\sigma_\dimens),
  \xi_1, \dotsc, \xi_d\right)\in \mathbb R^{2\dimens} .
$$

\noindent
Let $F_j^\bthres$ denote the $j^{th}$ marginal distribution
conditionally on $Y_j$ \emph{not} exceeding $\thres_j$, and let 
$\pexc_j =\mb
P(Y_j>\thres_j)$ denote the probability of excursion above $v_j$. 
The $j^{th}$  marginal model ($1\le j\le\dimens$) is thus 
\begin{equation}\label{marginal model}
\begin{aligned}
  F_j^{(\margpar)}(y) &= \prob(Y_{j,t} \le y\, | \,\xi_j, \sigma_j) \\
&=\begin{cases}
 1 - \pexc_j \left(1+\xi_j\frac{y-\thres_j}{\sigma_j}
\right)^{-1/\xi_j} &  (y\ge \thres_j),   \\
(1-\pexc_j) F_j^\bthres (y) & (y<  \thres_j).
\end{cases}
\end{aligned}
\end{equation}
It is common practice \citep{coles1991modeling,davison1990models} to
use  an empirical
estimate $\hat \Pexc= (\hat \pexc_1,\dotsc,\hat \pexc_d)$ for  the
vector of probabilities of marginal excursion, and to %an excess,  and to
ignore any estimation error, so that $\hat{ \Pexc}$ is identified to
$\Pexc$ is the sequel.

\subsection{Joint inference in a Poisson model}

When it comes to simultaneous estimation of the margins and of the
angular measure, the angular model  \eqref{eq:angularModel} for radial
excesses is difficult to handle, because a  radial failure region
$r>r_0$ on the Fréchet scale (\ie~, in terms of $\mb X$'s) corresponds to a
complicated shaped failure region on the original scale, which depends on
the marginal parameters and, accordingly, potentially   contains a varying number of
data points. It seems more reasonable to use a failure
region which is fixed on the original scale (in terms of $\mb Y$'s). Further,
 a common criticism towards radial failure regions
 \citep{ledford1996statistics} is that the marginal Pareto model is
 not valid near the axes of the positive orthant.  Last but not least,
 censoring occurs along the directions of the Cartesian coordinate
 system, which prevents  using  the polar model
 \eqref{eq:angularModel} as it is. 
To address these issues, the statistical model  for threshold excesses  developed in this
paper  uses a `rectangular' threshold.
Also, it will be very convenient (see
Section~\ref{sec:censorThres}) to adopt a 
Poisson process 
 representation of extremes \cite[see
 \emph{e.g.}][]{coles1991modeling} as an alternative to the POT 
 model \eqref{eq:angularModel},  with 
a  `censored likelihood' near the axes.  %  \cite[as \emph{e.g.} in
% ][ see section \ref{sec:censoredModel} below]{ledford1996statistics}.  
% As an alternative to the polar   excesses above large
% thresholds may be  modelled by a Poisson process
% ($\PRM$).

% Let us first introduce the Poisson model. Censoring issues are
% deferred to Section~\ref{sec:censoredModel}.
\paragraph{{\textbf{Poisson model}}}
  Under the same condition of domain of attraction as above, the  point
process formed by time-marked, standardized and suitable re-scaled data   converges
in distribution to a Poisson process %% holds 
 \citep[see \emph{e.g.}][]{resnick1987extreme, Resnick07,
  coles1991modeling}, 
% The DOA condition is equivalent to marginal
% convergence and

\noindent
\begin{equation}\label{cvPoisson}
\sum_{t=1}^{n} \mathds{1}_{(\frac{t}{n},\frac{\mb{X}_t}{ n})}
\overset{w}{\longrightarrow} \PRM( \leb\otimes\expmeas)\,,
\end{equation}

\noindent
 in the space of point measures on  $([0,1]\times \mb E)$, where $\mb
 E = [0,\infty]^d\setminus\{0\}$.
% Concerning the limiting intensity
%  measure, 
The temporal component  $\leb$ of the limiting intensity measure
denotes the
 Lebesgue measure on $\mathbb R$ and $\expmeas$, the so-called
 \emph{exponent measure},  % $\expmeas$  
 is homogeneous of order $-1$, and is related to the angular measure
 $H$ \emph{via} %  which is
% conveniently expressed in  pseudo-polar coordinates. 
% Consider  the $L_1$  norm on $\mathbb{R}^d$,  $\|\mb x\| = \sum_{i=1}^d
% |x_j|$. Then   the
% positive orthant of the 
% unit sphere  is the unit simplex 
%  $\simpl = \{
% (\ang_1,\dotsc,\ang_d ):\,\ang_j \ge 0\,,\; \sum_{j=1}^d \ang_j =
% 1\}\,.$ 
%  In this context, % the image measure of $\expmeas$ by the
% % polar transformation $\mb x 
% if $ (r,\Ang) =
%  (\|\mb x\|, \frac{\mb x}{ \|\mb x \|} )\in (0,\infty]\times \simpl$,
%  the exponent measure is a product measure 
 % in the pseudo-polar
 % coordinate system:
 % is a product
 % measure. Namely, identifying the image measure with $\expmeas$ for the
 % sake of conciseness, one  may write

\noindent
 \begin{equation}\label{productMeasure}
   \ud \expmeas(r,\mb w) = \frac{d}{r^2} \;\ud r\ud H(\mb w)\,.
 \end{equation}

% In practice,  temporal independence  may not hold.  For stationary sequences, if a short dependence condition
% of mixing type  is
% satisfied  \citep[condition $D$, see
% \emph{e.g.}][]{leadbetter1983extremes}, 
% declustering methods allow to treat cluster maxima as independent
% observations, see section \ref{sec:preproc} for details.

% If $n$ is the number of
% observed data,
From a statistical perspective, consider a  failure region $A_{\bthres}=\mb E\setminus
[\mb 0,\bthres] $, where $\bthres$ is the high multivariate threshold
introduced in section~\ref{sec:marginalModel} and  $[\mb 0,\bthres] =
[0,\thres_1]\times\dotsb\times[0,\thres_d]\setminus\{\mb 0 \}$. 
Call `excess above $\bthres$' any point $\mb Y_t$ in $A_\bthres$, as
opposed to marginal excesses %   and `marginal % excess' any 
$Y_{j,t}>\thres_j$.  
The Fréchet re-scaled multivariate threshold is 

\noindent
$$
\bfthres = \mb T(\bthres) = -1/\log(1-\Pexc)
$$ 

\noindent
and does not depend on $\margpar$. 
Consider the re-scaled  region on the Fréchet scale
$$ A_{\bfthres,n} = \frac{1}{n}\mb T (A_{\bthres} ) =
[0,\infty]^d \setminus
[0,\frac{\fthres_1}{n}]\times\dotsb\times[0,\frac{\fthres_d}{n}]\,,$$
and  denote  $A_{\bfthres}= A_{\bfthres,1}$. Applying the marginal transformations

\noindent
$$ \mathcal{T}_{j}^{ \margpar}(y) = -1/\log\left(F_j^{(\margpar)}(y)
\right )\,,$$ 

\noindent
the marginal variables $ X_{j,t} = \mathcal{T}_j^{\margpar}(Y_{j,t}) $
have  unit
Fréchet distribution, as %$\prob( X_{j,t}\le x) = \exp(-\frac{1}{x})$, as
required in \eqref{cvPoisson}.  The point process 
 $\mb N=  \sum_{t=1}^n \mathds{1}_{(\frac{t}{n},\frac{\mb X_t }{n})}$
% $(\frac{t}{n},\frac{\mb
%   X_t}{n})$'s such that
composed of the excesses $\mb X_t\in A_{\bfthres}$ (\ie $\mb Y_t\in
A_{\bfthres}$) is modeled according to the limit in
\eqref{cvPoisson}, % assumed to be the points of a Poisson process over
% on  $[0,1]\times A_{\bfthres,n}$, % with intensity measure
% %%$\ud s\times \ud \expmeas$
%  given by the right-hand side of \eqref{cvPoisson}.  The Poisson 
%  model for the point process  is thus 

\noindent
\begin{equation*}
%  \label{eq:poissonModel}
 \sum_{t=1}^n \mathds{1}_{(\frac{t}{n},\frac{\mb X_t %%%  T^{\margpar}( \mb Y_t )
}{n})}  \sim
  \PRM( \leb \otimes \expmeas) \text{ on } [0,1]\times A_{\bfthres,n}\,,
\end{equation*}

\noindent
where $\expmeas$ is of the
form \eqref{productMeasure}, with angular component $H$ written as a
Dirichlet mixture of the form \eqref{dirimixDens}. 

% {\color{blue}
% In the sequel, 
% $\mb T^{\margpar}$ denotes the vectorial transformation: 

% \noindent
% $$\mb T^{\margpar}(\mb y) = \left(\mathcal T^{\margpar}_1( y_1),
%   \dotsc, \mathcal T^{\margpar}_\dimens( y_\dimens)\right).$$

% % \noindent
% %  The marginal transformations above threshold have inverse Jacobian
% %  $J_j^{\margpar}$
% % $\mathcal{T}_j^{\margpar}: y_j\mapsto x_j $ above $\thres_j$, which is 

% % \begin{equation*}\label{jacobian}
% %  J_j^{\margpar}(y_j) =% \frac{1}{n}
% % \sigma_j^{-1}(\pexc_j )^{-\xi_j}
% % x_{j}^2 e^{\frac{1}{x_{j}}} \left[1-e^{\frac{-1}{x_{j}}}
% % \right]^{1+\xi_j} \,,
% % \end{equation*}
% %  where $x_j = \mathcal{T}_j^{\margpar}(y_j)$, which will appear in the likelihood contribution of marginal
% % excesses. 
% }

%   The Fréchet-rescaled multivariate threshold is 

% \noindent
% $$
% \bfthres = \mb T(\bthres) = -1/\log(1-\Pexc)
% $$ 

% \noindent
% and does not depend on $\margpar$.
% %Applying the marginal transformations

\paragraph{\textbf{Joint likelihood of  uncensored data}}
Let $\theta = \left(\margpar, \dmpar\right)$ be the parameter
for the joint model. As explained at  the beginning of this section, 
the model likelihood needs to be expressed in Cartesian coordinates.
%  Since  marginal transformations are applied in the
% Cartesian coordinate system, it is convenient to express the Poisson
% likelihood in Cartesian coordinates as well. In addition, censoring will
% involve integration along the Cartesian axes. 
% Thus, one needs the expression for
The density of an  exponent measure  
$\expmeas$ with respect to the $\dimens$-
 dimensional Lebesgue measure $\ud\mb x = \ud x_1\dotsb\ud x_d$,
  is  \citep[][Theorem~1]{coles1991modeling}

\noindent
\begin{equation}
  \label{eq:cartesianExp}
\frac{\ud\expmeas}{\ud\mb x} (\mb x)=   
\dimens\,.\, r^{-(d+1)} h(\Ang )\,.   
\end{equation}

\noindent
Denote by  $\lambda_\dmpar$ the exponent measure corresponding to the
Dirichlet mixture $h_\dmpar$.  % with parameter $\dmpar$. 
Then, 
 in the simplified case where the   $\mb Y_{j,t}$'s are exactly observed
  and where  the 
marginals $F_j$'s below threshold are known, the likelihood in the
Poisson model over $A_{\bthres}$ %  of the
% Poisson process
is

\begin{equation}\label{eq:fullLikelihood}
\begin{aligned}
  \mathcal{L}_{\bthres} \left( \{\mb y_t\}_{1\le t
\le n}, \theta
  \right) 
  & \propto e^{-n\,\expmeas_{\dmpar}(A_{\bfthres})}
  \prod_{i=1}^{n_{\bthres}} \Big\{
  \frac{\ud\expmeas_{\dmpar}}{\ud \mb x} ( \mb x_{t_i})%T^{\margpar}(\mb Y_{t_i}))
\prod_{ j: y_{j,t_i} > \thres_j} J_j^{\margpar}(y_{j,t_i})
  \Big\}\,,
\end{aligned}
\end{equation}

\noindent 
where  $t_1,\dotsc,t_{n_{\bthres}}$ are  the occurrence times of 
excesses  $\mb y_{t_i}\in A_{\bthres}$, 
$x_{j,t_i}~=~T_j^\margpar
%-1/\log\,F_j^{\margpar}
(y_{j,t_i})$,  and the
$J_j^{\margpar}$ are Jacobian terms resulting from marginal
transformations $T_j^\margpar$ (see Appendix~\ref{ap:simpleLikelihood}
for details).

\section{Censored model}\label{sec:censoredModel}
\subsection{Causes of censoring}
The presence of censored observations  is the result of two distinct causes:  first,  data are
partially observed, which results in interval- or
right-censoring, which we call \emph{natural censoring}.  
  In addition, observed data points that exceed at least
one threshold in one direction do not necessarily exceed all
thresholds, so that the marginal extreme value model does not apply.
Following \cite{ledford1996statistics}, those components are also
considered as 
left-censored. This second censoring process is thus  a
consequence of an 
inferential framework which is designed for analyzing extreme values
only, and we call it \emph{inferential censoring}.  

The total censoring process $\mathscr{C}$, which results from 
the juxtaposition of natural and inferential censoring,  is assumed to be non
informative. This means that  \citep[see also][]{gomez2004frequentist}, if $F$ is the marginal \emph{c.d.f.} for $Y_j$ and $f$ is the
marginal density, then $Y_j$'s  distribution conditional upon having
observed only the left and right censoring bounds $(L,R)$ is
$f(\point)/[F(R)-F(L)]$. This definition is easily extended to the
multivariate case by replacing $F(R)-F(L)$ by the integral of the
density over the  censored    directions 
 \citep[see \emph{e.g.}][for a proof of
consistency of maximum censored
likelihood estimators]{schnedler2005likelihood}.

\subsection{Natural censoring}
% In a first step, we ignore the uncertainty arising from potential
% measurement errors and tranformation of water levels into discharge. Even so, most of the $N = 14\,841$ daily data are missing or
% censored (for
% example, during the historical period, on days where no event was
% recorded).

Call `Natural censoring' the one  which occurs %`natural censoring' which occurs
independently from the choice of an extreme threshold $\bthres$ by the
statistician. 
The observed process is
denoted 
${\mb O} = ({\mb O}_t)_t$,  with $\mb O_t=(O_{1,t},\dotsc,O_{d,t})$.
One marginal observation $O_{j,t}$ consists in a label $\kappa_{j,t}$
indicating   presence or absence of censoring, together with   the exact data
$Y_{j,t}$ (if observed)  or the censoring
bounds $(L_{j,t}, R_{j,t})$ (where $L_{j,t}$ may be set to $0$ in the
case of left-censoring or missing data and $L_{j,t} = +\infty$ in the
case of right-censored or missing data). 
In the sequel, $\kappa_{j,t}= 0$ (\emph{resp.} $ 1,2,3$)  refer respectively to missing,
exact, right- and left- censored data. 

\begin{figure}[hbtp]
\centering
\makebox{\includegraphics[scale=0.3]{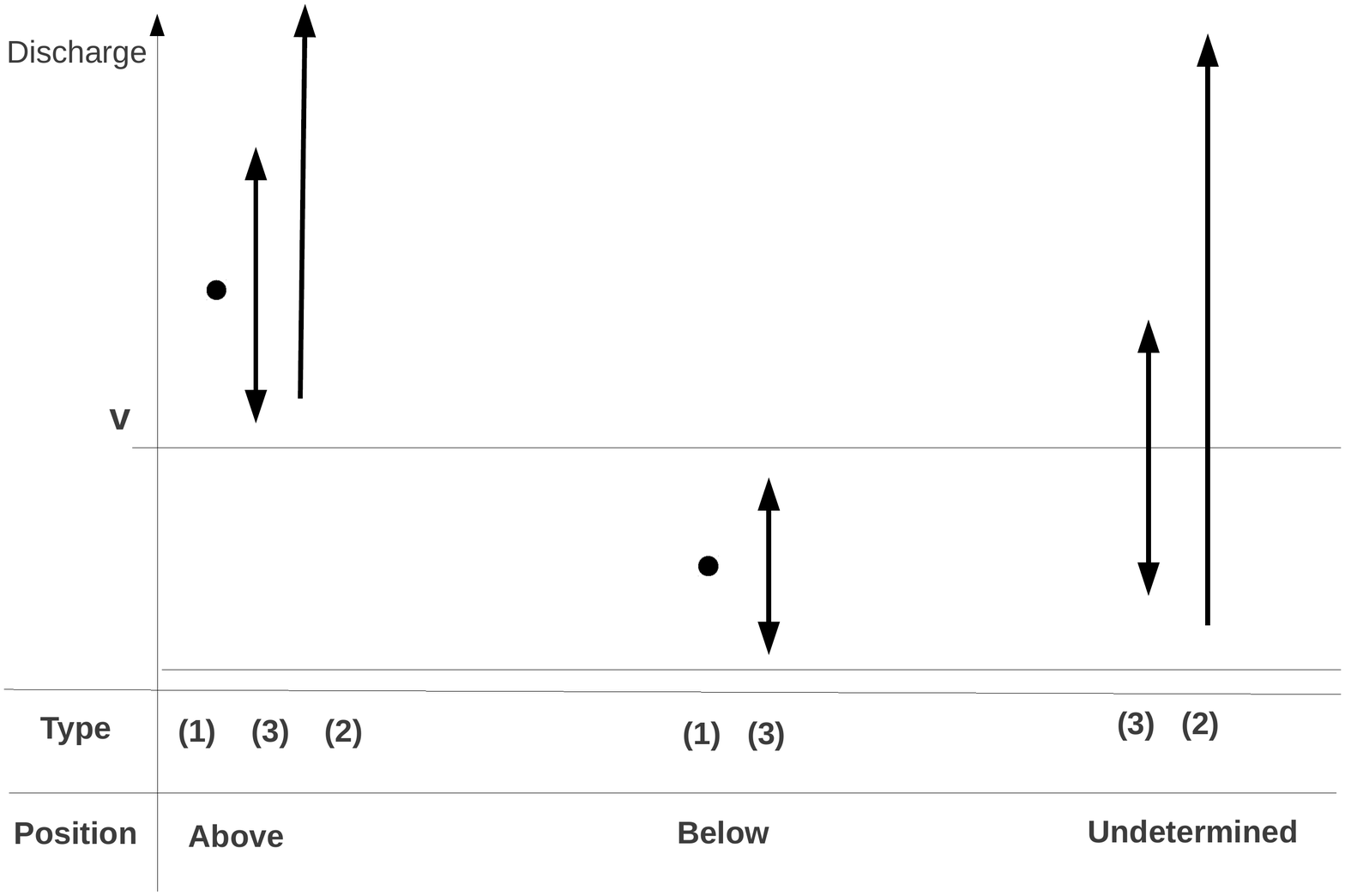}}
\caption{\label{fig: excesses}Position  of marginal data points   with respect to a
  marginal threshold $\thres$ (horizontal line). Black dots:  marginal data points  of
  type $\datatype=1$; vertical arrows: data of  type $\datatype\in
  \{2,3\}$. 
  }
\end{figure}

In this context,   the `position' of a marginal data $O_{j,t}$ with respect to
the threshold is not necessarily well defined for censored data.
Recall that we consider a case where the process of interest $(\mb
Y_t)_{t\ge 0}$ is stationary, whereas  the censoring bounds vary
with time, as a result of external factors on the observation process.
 When the censoring interval overlaps the
threshold, \ie,  $\thres_j\in (L_{j,t},R_{j,t}), \datatype_j\in\{2,3\}$), the statistician
does not know if an excess occurred or not. This situation is
described here  as \emph{$O_{j,t}$ marginally overlapping the threshold}.
The different positions of $O_{j,t}$  with respect to
$\thres_j$ encountered in the data set of interest in this
paper are  summarized in Figure~\ref{fig: excesses}.

% In this context, the definition of the position of a marginal data
% relatively to a given threshold $\thres_j$ requires some care.  The
% different situations are summarised in Figure~\ref{fig: excesses}.
% $O_j$ exceeds $\thres_j$ (left panel) if there is no censoring and %
%                                                                    % $\datatype_j=1$
%                                                                    % and
% $Y_j > \thres_j$, or if %$\datatype_j\in\{2,3\}$ and
% $L_j>\thres_j$. Similarly, $O_j$ is below $\thres_j$ if
% $\datatype_j=1$ and $Y_j\le \thres_j$, or if $\datatype_j=3$ and $R_j\le
% \thres_j$. If none of the above conditions hold, \emph{i.e.} when censoring occurs
% with censoring interval intersecting the threshold, the relative
% positions of $Y_j$ and $\thres_j$ are
% unknown, and we say that $O_j$ has \emph{undetermined position} with
% respect to $\thres_j$.

For a  multivariate observation $\mb O_t$,  if at least one coordinate
marginally overlaps the threshold or is   missing, and if the others are below
the threshold, then the position of $\mb O_t$  with respect to the
\emph{multivariate threshold} $\bthres$ is undetermined. Indeed, 
$\mb O_t$ is \emph{below threshold} if all its marginals are below the
corresponding marginal threshold, and \emph{above threshold}  (in
the failure region)
otherwise. In the undetermined case, $\mb O_t$ is qualified  as  \emph{globally
  overlapping the threshold}.
% has undetermined position with
% respect to $\bthres$.

% A marginal data  $O_{j}$ is said to exceed  $\thres_j$  if at
% least  one
% of the following conditions is  fulfilled (see 
% \begin{itemize}
% \item $ \datatype_{j} = 1$ and $ Y_{j} \ge \thres_j$,
% \item $ \datatype_{j} \in \{ 2,3\}$ and $ L_{j} \ge \thres_j$.
% \end{itemize}
% Similarly, $O_{j}$ is below $\thres_j$ ($O_{j} \prec \thres_j$) if one of the
% following holds: 
% \begin{itemize}
% \item $ \datatype_{j} = 1$ and $ Y_{j}  < \thres_j$,
% \item $ \datatype_{j} = 3 $ and $ R_{j} < \thres_j$.
% \end{itemize}
% % If neither $O_{j} \prec \thres_j$ nor $O_{j} \succ \thres_j$
% % (\emph{e.g.} when $ \datatype_{j} = 3$ and $ L_{j} \le \thres_j \le  R_{j}$
% % ),

% %% the data coordinate is considered as missing. If 

% If none of the above cases holds, but $\datatype_{j}\neq0$, then $O_{j}$ is said to be `undetermined' with respect to
% $\thres_j$.  This is occurs when 
% \begin{itemize}
% \item $\datatype_{{j}} = 2  $ and $L_{j}<\thres_j$, or
% \item $\datatype_{j} = 3$,  $L_{j}< \thres_j$ and $R_{j}> \thres_j$.
% \end{itemize}

\subsection{Inferential censoring below threshold}%Censoring observations below threshold}
\begin{figure}[hbtp]
\centering
\makebox{\includegraphics[scale=0.3]{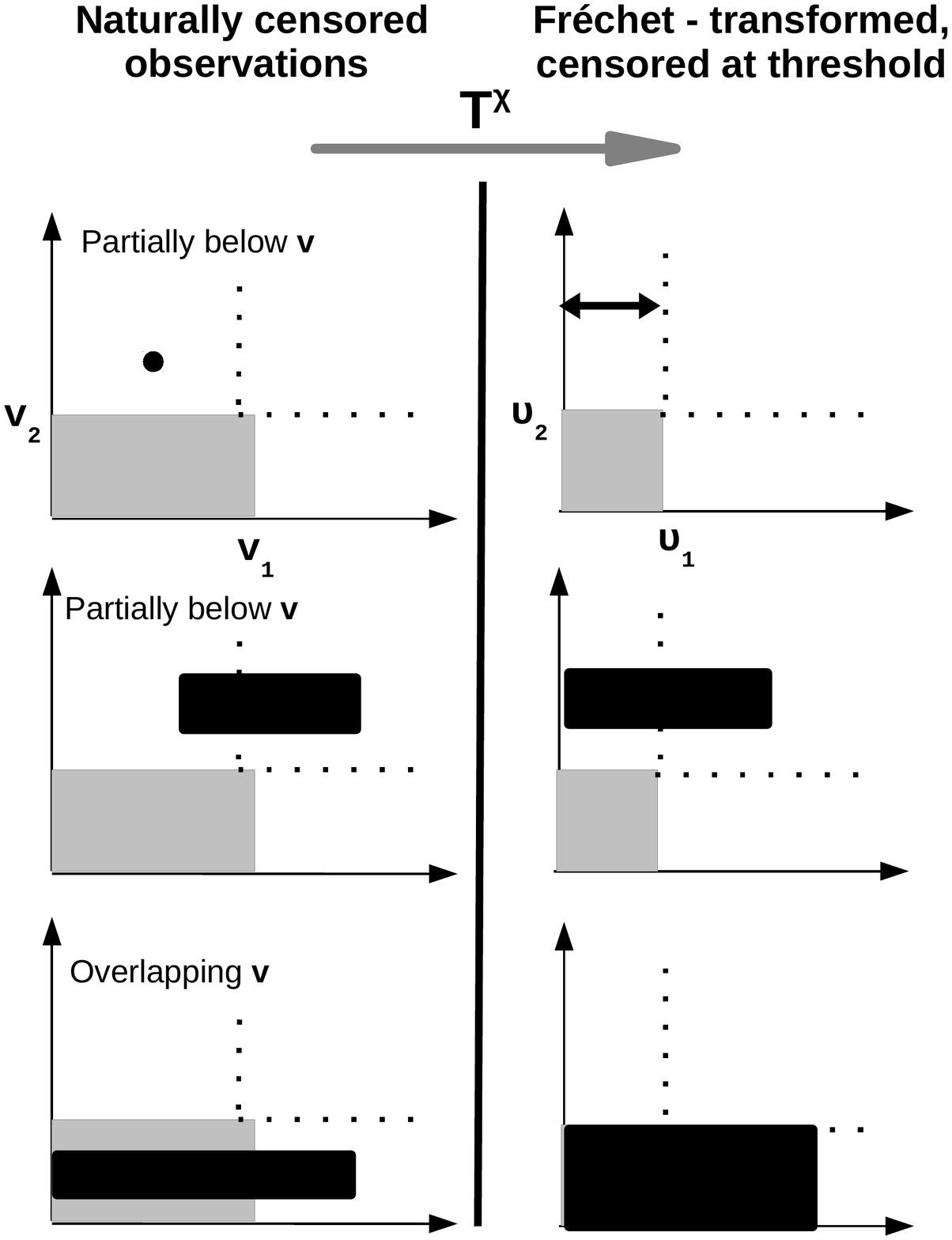}}
\caption{\label{fig:censorTransform} Example in the two-dimensional case  of marginal transformation and censoring below
  threshold, for data of marginal types $(1,1)$, (upper panel)
  $(3,1)$ (middle
  panel) and $(3,3)$ (lower panel). 
In these three cases,  observations  are  represented in black,
respectively by  a
dot, an horizontal arrow and a rectangle. The Grey areas represent
respectively the
multivariate threshold $\bthres$ (left side) and  the
Fréchet transformed one $\bfthres$ (right side). The two upper  panels
correspond to observations above threshold, while the lower panel
shows an undetermined observation.}
\end{figure}

\label{sec:censorThres}
Since the marginal distributions $F_j^{\bthres}$'s, conditional upon
not exceeding $\thres_j$,  are unknown, the $X_{j,t}$'s such
 that $Y_{j,t} <\thres_j$ are not available. Instead of attempting to estimate the
$F_j^{\bthres}$'s, one option  is  to censor the Fréchet-transformed components below
threshold.
More precisely,  for a raw observation $O_{j,t} = (\datatype_{j,t}, Y_{j,t}, L_{j,t}, R_{j,t} )$,
 let us  denote
  by $ 
C_{j,t}^{\margpar}  = (\tilde \datatype_{j,t}, X_{j,t}, \tilde
L_{j,t}, \tilde R_{j,t} )$   
 the corresponding `Fréchet transformed' and censored one,  and $\mb
C_t^\margpar$ the multivariate observation $(C_{1,t}^\margpar,\dotsc,C_{d,t}^\margpar)$.
 The  transformation $\mb O_{t} \mapsto \mb C_{t}^\margpar$
 is illustrated in  Figure~\ref{fig:censorTransform} in the bi-variate
 case. % ; 
%  a formal
% definition  is given  in appendix~\ref{ap:
%   formalCensFrech}. 

In a nutshell, 
inferential censoring  occurs when the censoring intervals
 are below threshold or  marginally overlapping it, or when an exact data
component $Y_{j,t} < \thres_{j}$ is recorded. 
A formal definition of $C_{j}^\margpar$ is as follows: 

\begin{itemize}
\item $(\tilde \datatype_{j,t},  X_{j,t}) = 
  \begin{cases}
    (3, \texttt{NA})& \text{if } \datatype_{j,t} = 1 \text{ and }
    Y_{j,t}<\thres_j\,, \\
    (0, \texttt{NA}) &\text{if } \datatype_{j,t} = 2 \text{ and }
    L_{j,t}<\thres_j\,, \\
    \big(\datatype_{j,t}, \mathcal{T}_j^\margpar(Y_{j,t})\big) & \text{otherwise}.
  \end{cases}
  $

\item $\tilde L_{j,t} = 
  \begin{cases}
    0 & \text{if } L_{j,t} < \thres_j % \text{ and } L_{j,t} \neq    \texttt{NA}\,, 
    \,,\\
    \mathcal{T}_{j}^{\margpar}(L_{j,t}) & \text{otherwise. }
  \end{cases} 
  $

\item $\tilde R_{j,t} =
  \begin{cases}
    \fthres_j  & \text{if } R_{j,t} < \thres_j \,,\\%%\text{ and }R_{j,t} \neq \texttt{NA}\,,\\
    \mathcal{T}_{j}^{\margpar}(R_{j,t}) & \text{otherwise.} 
  \end{cases} 
  $
\end{itemize}
In the above definition, $\texttt{NA}$ stands for a missing %  or
% irrelevant
value and it is understood that 
$\mathcal{T}_j^{\margpar}(\texttt{NA}) = \texttt{NA}$.

 In the end, observations globally overlapping threshold have their
 marginal lower bounds $\tilde L_{j,t}$ set to zero  if $L_{j,t}<v_j$. 
The interest of using a Poisson model  becomes clear at this
point. Indeed, 
censored observations $\mb C_{t}^\margpar$ obtained from observations $\mb
O_t$ globally overlapping threshold correspond to events 
of the kind  `\emph{The observation at time $t$ belongs to 
   $\mb{[0},\mb{\tilde R}_t\mb{]}$}', which, by contrapositive, means `\emph{No point
is observed  outside of $\mb{[0},\mb{\tilde R}_t\mb{]}$ between $t$ and
$t+1$}'.  This  is written in terms of the Poisson process $\mb N$ as  % events of the kind

\begin{equation}
  \label{eq:missingFailurSet}
\mb N\left(\left[\frac{t}{n},\frac{t+1}{n}\right)\times \left[\mb 0,
   \frac{\mb{\tilde 
     R}_t}{n} \right]^c\right) = 0\,,
\end{equation}
 
\noindent which is a 
measurable event with respect to $\mb N$. %  namely 
 % % outisde belongs to $[\frac{t}{n},\frac{t+1}{n}]\times
 % %  \frac{1}{n}\mb{[0},\mb{R}_t\mb{]}$},
 % which 
%\noindent
 % where $N$ is the considered 
 % point process.
The overlapping observations thus have a well defined likelihood in a
Poisson model, as detailed in the next section, whereas they could not
be taken into account in \cite{sabourinNaveau2012}'s POT
model.  

\subsection{Poisson likelihood with  censored and missing data}
\label{subsec:adaptCensor}

% In polar coordinates, if $H$ is a \DM\ spectral measure with parameter
% $\dmpar$, the exponent measure   in \eqref{productMeasure} is 
% \begin{equation}
%   \label{eq:poissonDMPolar}
%   \ud \expmeas_{\dmpar} = \frac{d}{r^2} h_{\dmpar}(\Ang)\ud r\ud\Ang\,.
% \end{equation}

% In  presence of censored or missing components, the likelihood of the
% Poisson process involves 
% partial 
% integration of the exponent measure $\expmeas$  along  the axes of the Cartesian coordinate system, 
%  the integration being performed in the
% direction of the missing or censored  coordinates.

Due to the combination of natural and inferential censoring, the
data set (from which missing days are excluded)  is decomposed into  data in the failure
region,  data overlapping threshold and data below threshold. Let
$n_{\bthres}$, $n_{\bthres}'$ and $n_{\bthres}''$ be the respective
number of observations in each category. % , so that
% Finally,%  let  $n''_{\bthres}$ be 
% % the number of observations below threshold.
% % T
The number of non missing days is thus 

\noindent
$$
n_{\text{obs}} =n_{\bthres} + n'_{\bthres} + n''_{\bthres}\,,$$

\noindent
 and the number of `determined' data (\ie not overlapping $\bthres$)
is 

\noindent
$$\ndef = n_{\bthres}+ n''_{\bthres}\,.$$

\noindent
The  $n_\bthres$ Fréchet-transformed  observations  $\{\mb {C}_{t_i}^{\margpar}\} (i\in
\{1,\dotsc,n_{\bthres} \})$   correspond to events of the kind 

\noindent
$$
\mb X_{t_i} \in \mb [\mb{\tilde L}_{t_i} \boldsymbol{, \tilde
  R}_{t_i} \mb ] \qquad\text{(rectangular region)}
$$

\noindent where $\tilde R_{j,t_i} = \tilde L_{j,t_i} = X_{j,t_i}$ in
the case of 
exact data.  
% are denoted 
% $\mb{C}^{\margpar} = \{\mb {C}_{t_i}^{\margpar}\} (i\in
% \{1,\dotsc,n_{\bthres} \}) $, where ${C}_{t_i}^{\margpar}= ({C}_{1, t_i}^{\margpar},\dotsc,{C}_{d, t_i}^{\margpar})$ has been
% defined in section~\ref{sec:censorThres}.

 Observations overlapping threshold correspond to events of the kind
 \eqref{eq:missingFailurSet} introduced in the previous section. 
When a limited number $\nblock $  of right censoring bounds
 $\mb{ R}_t=(R_{1,t},\dotsc, R_{d,t})$
 are
present,  it is convenient to classify these overlapping events 
% can be further classified 
accordingly, 
% their right censoring bound $\mb{\tilde R}_t=(R_{1,t},\dotsc,
% R_{d,t})$ 
 writing  %   so that we may write 
 $n'_{\bthres} = \sum_{i=1}^{\nblock} n'_{i}$
  where 
%$\nblock$ is the number of different right censoring bounds, 
  $n'_{i} $ is the
number of   observations with right censoring bound $\mb{R}_{t_i}$. 
Since the Poisson
process is temporally exchangeable, there is no loss of generality in assuming
that  the latter observations  occur at  consecutive dates 
$(t'_i,\dotsc, t'_i+n'_i-1)$.  
The  region  $\mb
E \setminus[\mb 0,\mb{\tilde  R}_{t'_i}]$ is `missed' by the
Fréchet re-scaled process  $\mb X_t$ during this time
period.%  On the Fréchet scale, this corresponds to a `missed' region 

% \noindent
% $$
% A'_{i} := \mb
% E \setminus[\mb 0,\mb{\tilde{ R}}_{t'_i}] \,.
% $$
 
\noindent
With theses notations, the censored likelihood in the Poisson model may be written % involves 
% partial 
% integration of the exponent measure $\expmeas$  along  the axes of the Cartesian coordinate system, 
%  the integration being performed in the
% direction of the missing or censored  coordinates.

\noindent
\begin{equation}\label{censoredLkl_full}
\begin{aligned}
  \mathcal{L}_{\bthres}(\mb O, \theta)
&=\exp\Big[- \ndef\,\expmeas_{\dmpar}(A_{\bfthres} ) -
\sum_{i=1}^{\nblock} n'_{i}
\expmeas_{\dmpar}(A'_i)\Big]%\big([\mb 0,\tilde {\mb{R}}_{t'_i}]^c \big) \Big]
 \times\dotsb\\
&\dotsb\,\prod_{i=1}^{n_{\bthres}} \Big\{
  \int_{\mb [\mb{\tilde L}_{t_i} \boldsymbol{, \tilde
  R}_{t_i} \mb ]}
%%{ \mathscr{C}_{t_i}^\margpar(\mb x) \in \mb {C}_{t_i}^{\margpar} }
%\left[\mathscr{C}_{t_i}^\margpar \right]^{-1}( \mb {C}_{t_i}^{\margpar})}
%{\mathscr{E}_{t_i}^\margpar(\mb C_{t_i}^{\margpar}) }
  \frac{\ud \expmeas_{\dmpar}}{\ud \mb x} \ud\ell_i(\mb x)
  \prod_{ j: y_{j,t_i} > \thres_j} J_j^{\margpar}(y_{j,t_i}) \Big\}\,,
\end{aligned}
\end{equation}

\noindent
where notation `$\ud \ell_i(\mb x)$' in the integral terms is a shorthand
for `the Lebesgue measure of dimension equal to that of $\mb [\mb{\tilde L}_{t_i} \mb{, \tilde
  R}_{t_i} \mb ]$' when the latter is greater than one, or `the Dirac
mass at $\mb x_{t_i} = \mb{\tilde L}_{t_i} = \mb{\tilde R}_{t_i} $' for exact data. 
% \noindent
% where  $(\mathscr{C}_t^\margpar)_t$
% %  =
% % (\mathscr{C}_{t_i}^\margpar)_{1\le i\le n_{\bthres}},
% % (\mathscr{C}_{t'_i}^\margpar)_{1\le i\le n'_{\bthres}}$ 
% is the censoring
% process on the Fréchet scale, which transforms
% Fréchet points $\mb x_t$ in  $A_{\bfthres}$ % The exponential terms in  \eqref{censoredLkl} have been   obtained as a
% modification of those in 
Compared with the uncensored likelihood  \eqref{eq:fullLikelihood},
  $n$  has been  replaced with $n_{\text{obs}}$,  
%$\ndef$ 
  the exponential term  for the non overlapping   data follows from

\noindent
$$
\exp\Big(- 
\frac{\ndef}{n_{\text{obs}}}
\expmeas_{\dmpar}\big(A_{\bfthres, n_{\text{obs}}} \big)\Big)=%\ndef}) = 
\exp\Big(-\ndef\, \expmeas_{\dmpar}\big(A_{\bfthres}\big)\Big)\,,
$$

\noindent
and  a similar argument yields the  additional terms 
$\exp(- n'_{i}
\expmeas_{\dmpar}(A'_i)) $  for overlapping  data.%  are obtained
% similarly. % comes from the fact that 

% %\noindent
% $$
% \begin{aligned}
% \mb P_{\theta}\left\{N\left( \left[\frac{t'_{i}}{n_{\text{obs}}},
% \frac{t'_i+ n'_{i} -1}{n_{\text{obs}}} \right]
% \times \frac{1}{n_{\text{obs}}}A'_{i}\right) %%\mb{\Big[0},\frac{1}{n_{\text{obs}}}\mb{ \tilde  R}_{t'_i}\Big]^c \right) 
% = 0\right\} % &= 
% % \exp\left(
% % -\frac{n'_{i} }{n_{\text{obs}} }
% % \expmeas_{\dmpar}\Big( A'_{i,n_{\text{obs}}}\Big)
% % %mb{\big[0},\frac{1}{n_{\text{obs}}}\mb{\tilde R}_{t'_i}\big]^c \Big)
% % \right) \\
% &= 
% \exp\left(
% - n'_{i}\, 
% \expmeas_{\dmpar}\big(A'_i\big)
% \right)\, .
% \end{aligned} 
% $$

At this stage, the model has been entirely specified. The remaining
issue concerns the treatment of the  integral terms 

\noindent
\begin{equation}
  \label{eq:integrTerms}
\int_{\mb [\mb{\tilde L}_{t_i} \mb{, \tilde
  R}_{t_i} \mb ]}
%%{ \mathscr{C}_{t_i}^\margpar(\mb x) \in \mb {C}_{t_i}^{\margpar} }
%\left[\mathscr{C}_{t_i}^\margpar \right]^{-1}( \mb {C}_{t_i}^{\margpar})}
%{\mathscr{E}_{t_i}^\margpar(\mb C_{t_i}^{\margpar}) }
  \frac{\ud \expmeas_{\dmpar}}{\ud \mb x} \ud\ell_i(\mb x)
\end{equation}

\noindent
and the exponential terms 

\noindent
\begin{equation}
  \label{eq:expTerms}
\exp\Big[- \ndef\,\expmeas_{\dmpar}(A_{\bfthres} ) \big] 
 \qquad  \text{and} \qquad 
\exp\Big[- n'_{i}
\expmeas_{\dmpar}(A'_i)\Big] \,,  
\end{equation}

\noindent which have no analytic expression, as they require
integrating $\lambda_\dmpar$ over rectangular regions.  First, the
dimension of numerical integration can be reduced as far as `missing
coordinates' are involved, because partial integration of
$\lambda_\dmpar$ over $[0,\infty]$ in one direction has an exact
expression. The model is stable under marginalization, in the sense
that the obtained marginal exponent measures correspond again to Dirichlet
mixtures on a lower dimensional simplex (see Appendix~\ref{sec:integrMissing} for
details).  However, no closed form is available for the integral in
the remaining censored directions, nor for the exponent measures of
$A_{\bfthres}$ or the $A'_i$'s. This problem is tackled in the next
section \emph{via } a data augmentation method.

% \section{Algorithm for inference based on a censored likelihood}%-based inference}% : Data augmenting and MCMC
%   % algorithm}
% \label{sec:censoredInf}
\section{Data augmentation}\label{sec:data_aug}

\subsection{Background}
In a Bayesian context, one major objective is to generate parameter samples
approximately distributed according to the posterior. 
 In classical \MCMC\ algorithms, the value of  the likelihood is
 needed to define the transition kernel. Evaluating  the integrated
likelihood $\mathcal{L}_{\bthres} (\mb O, \theta)$ % (\emph{e.g.} 
% by a Gaussian quadrature method)
% or a Monte-Carlo integration) 
at each
 iteration  of the algorithm seems unmanageable: % would dramatically slow down the
% execution:
The
dimension of integration can grow up to $\dimens$, for each observation, and the shape of the integrand varies from one iteration to
another, which is not favorable to standard quadrature methods. 
In particular, large or low values of the shape parameters $ \nu_{m}$ in
$\dmpar$ induce concentration of the integrand around the centers
$\Mu_{m}$ or  unboundedness at the simplex boundaries. 
%  Another technical issue arises from the terms
% $e^{-\ndef\,
% \expmeas_{\dmpar}(A_{\bfthres})},\,
%  e^{-n'_i\,\expmeas_{\dmpar}(A'_i)} $ 
%  in \eqref{eq:fullLikelihood}, which have no
% analytic expression.
% Both problems can be addressed with 
Instead, data augmentation methods \citep[see
\emph{e.g.}][]{tanner1987calculation, van2001art} treat missing or
partially observed data as additional parameters, so that the
numerical integration step is traded against an increased dimension of
the parameter space.  In this section, $[\point]$ denotes the
distribution of a random quantity as well as its density with respect
to some appropriate reference measure.  Proportionality between
$\sigma$-finite measures is denoted by $\propto$.  Thus, $[\theta]$ is
the prior density and $[\theta\,|\, \mb O]\propto [\theta]\,
\mathcal{L}_{\bthres}(\mb O,\theta)$ is the posterior.  

The main idea is to define an augmentation space $\mathcal{Z}$, and a
probability measure $ [\point|\mb O]\aug$, on the augmented space
$\Theta\times \mathcal{Z}$, conditional on the observations, which may
be sampled using classical \MCMC\ methods, and which is consistent
with the `objective distribution' on $\Theta$. That is, the posterior on $\Theta$
must be obtained by marginalization of $[\point | \, \mb
O]\aug$,% up to a
% multiplicative constant

\begin{equation}
  \label{eq:consistencyAugmented}
  [\theta|\mb O] = \int_{\mathcal{Z}}[ z,\theta\, | \,\mb O]\aug\ud  z\,.
\end{equation}

In the sequel, the invariant measure (or its density) $[z,\theta\,|\,\mb
O\,]\aug$ is referred to as the \emph{augmented posterior}.
\subsection{Data augmentation in the Poisson model} 
\label{sec:DataAugmentPoissonModel}

In our context, finding an augmentation random variable $Z_+$ with
easily manageable conditional distributions, such that the augmented
posterior  $[ z,\theta |\,\mb O]_+$  satisfy
\eqref{eq:consistencyAugmented}, is far from straightforward, mainly
due to the exponential terms \eqref{eq:expTerms} in the likelihood.  

 Instead, an
 intermediate  variable $\mb Z$ is introduced, which plays the role
 of a proposal distribution in the MCMC algorithm. $\mb Z$ is defined
 conditionally to $\fullpar$, through the  \emph{augmented likelihood} 
%
%\noindent
$
[\,\mb z,\mb O\,|\,\fullpar\,] , 
$   
%
%\noindent
 so that the full conditionals 
$[\mb Z \; |\; \theta, \mb O]$ %and $[\theta\;| \;\mb Z, \mb O]$ 
can be  
 directly simulated as block proposals in a
 \emph{Metropolis-within-Gibbs}  algorithm \citep{tierney1994markov}. To ensure the
 marginalization condition \eqref{eq:consistencyAugmented},  the
 augmented  posterior
$[\mb Z,\theta|\, \mb O]\aug$ % which is  invariant under 
 % the Markov kernel
 is  \emph{not} the same as $[\mb Z,\theta | \,  \mb O]$. Instead it
 has a density of the form % but %  $[\mb
 % % z,\theta\;|\;\mb O]\aug$, which
 % takes the  form

 \begin{equation}
   \label{eq:InveriantMeas}
[\mb z, \theta\;|\;\mb O]\aug \; \propto \;  [\mb z ,\mb
O \;  |\;  \theta \,]\; [\,\theta\,]\; \varphi(\mb z)\,, %%\mathcal{L}'(\mb O,\theta)\,.   
 \end{equation}

 \noindent 
where  $\varphi(\mb z)$ is any
 weight  function allowing to enforce the consistency condition~\eqref{eq:consistencyAugmented}. %s 
    %%and $\mathcal{L}'(\mb O, \theta)$ 
 % is defined
 % so that
The $[\theta]$ terms cancel out and  the latter condition  is equivalent to 

\begin{equation}
  \label{eq:consistencyLikelihood}
\mathcal{L}_{\bthres} (\mb O, \theta) \propto 
% \left\{ 
\int[\mb z,  \mb O \; | \; \theta]\; \varphi(\mb z)\ud \mb z\,.
%\right\}\,.%%\mathcal{L}'(\mb O,\theta)\,.  
\end{equation}

\noindent
  In the end, a posterior sample
 from $ [\theta |  \mb O]$ is simply obtained by ignoring
the $\mb{Z}$-components from the one  produced  with the
`augmented' Markov chain.

In our case,   the augmented data  $\mb Z$ consist 
 of two parts, $\mb{Z} = (\mb Z_{\text{above}}, \mb Z' )$. The first
 one, 
  $\mb Z_{\text{above}} =\{\mb Z_{t_i}\}_{i \le
  n_{\bthres}}$,    accounts for integral terms
 \eqref{eq:integrTerms}, while the second one,  $\mb Z' = (\mb
 Z'_{\bfthres},\{ \mb Z'_i\}_{ i\le \nblock} )$, accounts    for the exponential terms
 \eqref{eq:expTerms}. 
The  $\mb Z_{t_i}$'s  have an intuitive interpretation, which is
standard in data augmenting methods: they  replace the censored
$X_{j,t_i}$'s. On the contrary, 
the $\mb Z'_i$'s and the $\mb Z'_{\bfthres}$  are just a computational trick accounting for the
exponential terms: they are  the points of independent Poisson
processes defined on `nice' radial sets   containing the failure
regions of interest $A'_i$'s and  $A_{\bfthres}$, and  $\varphi$ is a
smoothed version of an indicator function of the failure regions.% , 
% see Appendix~\ref{sec:dataAugmentDetails} and~\ref{consistencyPoisson}
% for details. 

% chosen in such a way that $\PE(\varphi(Z')) $ is the 

%  which 
%  account for the exponential terms %avoid the computation of 
% $e^{-\ndef\,\expmeas_{\dmpar}(A_{\bfthres})}$ and
%  $e^{-n'_{i}\,\expmeas_{\dmpar}(A'_i)}$, $(i\le \nblock)$.
The augmented likelihood  factorizes as 

\begin{equation}
  \label{eq:augmentedLKL}
[\;\mb z, \mb O\;|\;\fullpar\;] =\prod_{i=1}^{n_{\bthres}}\big\{[\,\mb z_{t_i}, \mb
O_{t_i}\;|\;\fullpar\,]\big\}\,
[\,\mb z'_{\bfthres}\;|\;\dmpar] \,\prod_{i=1}^{\nblock} [\mb z'_i\,|\,\dmpar]\,,  
\end{equation}

\noindent
and  the weight function   $\varphi$   is  of
the form 

$$\varphi(\mb z) = \varphi_{{\bfthres}}(\mb z'_{\bfthres})
\prod_{i=1}^{\nblock} \varphi'_{{i}}(\mb z'_{i})\,.
$$

\noindent A precise definition of $\mb Z_{\text{above}}$ is given in
Appendix~\ref{sec:dataAugmentDetails}, together with the expression of
the corresponding contributions to the augmented likelihood, $[\,\mb
z_{t_i}\,\mb O_{t_i}\,|\fullpar\,]$  .
The
full conditionals $[\mb Z_{\text{above}}  \,| \mb O\,,\fullpar\,]$  are derived
in \ref{ap: augmentedDistr}. The augmentation Poisson processes
 $\mb Z'$,
together with the weight $\varphi$ are defined in Appendix~\ref{sec:augmentExp},    and
the 
compatibility condition \eqref{eq:consistencyLikelihood}  is proved to
hold in Appendix~\ref{consistencyPoisson}.

\subsection{Implementation of a MCMC algorithm on the augmented space} \label{sec:MCMCalgo}

This section describes only the main features of the algorithm,
% implemented in this work in order  to
% estimate the posterior distribution $[\,\theta\;|\;\mb O]$,
 more
details are provided  in Appendix~\ref{sec:MCMC algo}. 
The present algorithm is an extension of  % is an adaptation of the one proposed by 
 \cite{sabourinNaveau2012}'s one, who proposed a  \emph{Metropolis-within-Gibbs}
 algorithm to sample the posterior distribution of the angular measure
 in a  POT framework \eqref{eq:angularModel}, within the Dirichlet
 mixture model \eqref{dirimixDens}.
The number of components in the Dirichlet mixture is not
fixed in their model, and the MCMC allows \emph{reversible-jumps}
between parameters sub-spaces of fixed dimension, each corresponding
to a fixed  number of components in the Dirichlet mixture. 
 Their algorithm
 approximates the posterior distribution $[\dmpar|\mathcal{W}]$,
 where $\dmpar$ is the DM  parameter and 
 $\mathcal{W}$ is an angular data set consisting of the angular
 components $W_t$ of  data $X_t$ that have been normalized to Fréchet
 margins in a preliminary step.  

% with reversible jumps  \citep{green1995reversible} between
%  sub-models with a given  number of Dirichlet mixture components. 
In contrast, the present  algorithm  handles `raw' data
(not standardized to Fréchet margins), with censoring of various types
as described in Section~\ref{sec:censoredModel}, so as to approach the
full posterior distribution (marginal and dependence parameters)
$[\,\theta\;|\;\mb O\,]$ . In practice, this
amounts to allowing two additional move types  in the
\emph{Metropolis-within-Gibbs} sampler: \emph{marginal moves} 
(updating the marginal parameters) and \emph{augmentation moves}
updating the augmentation data $\mb Z$ described above. The
 reversible jumps  and the moves  updating the DM parameters
 keeping the dimension constant 
 are unchanged.

\section{Simulations and real case example}\label{sec:results}

% This section illustrates the methodology with simulated data and the
% real case hydrological dataset mentioned in introduction. 
% The purpose   is not to study in full generality  model features such as 
%  \emph{e.g} the prior
% influence on the 
% estimates, or the convergence on the estimates with the sample size,
% which are  likely to be quantitatively most dependent on the censoring process. 
%  Instead, k
 Keeping
 in mind the application, the aim of this section is to 
verify that the algorithm  provides reasonable estimates with  data
sets that  `resemble' the particular one  motivating this work. 

After a  description of the experimental setting, an example of
results obtained  with a single  random data set is given, then a more
systematic study is conducted over 50 independent data sets. 
Finally, a brief description of the results obtained with the original
hydrological data  is provided. The latter part is kept short because, as
mentioned in introduction, from a hydrological point of view, a full discussion of the benefits brought
by 
historical information, as well as treatment of temporal dependence in
the case of heavily censored data  is needed  and will be
 the subject of  a separate paper. 

\subsection{Experimental setting}\label{sec:expeSetting}
For this simulation study, the marginal shape parameters are constrained to be
equal to each other, $\xi_1 = \dotsb = \xi_\dimens :=\xi$, in accordance
 with  the   \emph{regional frequency analysis} hydrological
framework \citep{hosking2005regional}, where the different gauging stations under study are
relatively close to each other (in the same watershed). Also, the
dimension is set to $\dimens=4$, as it is for the hydrological data
set of interest.

 Preliminary  likelihood maximization (with respect to the marginal
 parameters) is   performed on the 
 hydrological data, again  assuming 
independence between locations for the sake of simplicity and imposing a common shape parameter
(this latter  hypothesis not being rejected by a likelihood ratio
test). Then, data sets   are   simulated according to  the
 model for
excesses \eqref{eq:angularModel},
with marginal parameters and threshold excess probabilities (for
daily data) 
approximately equal to the inferred ones, \emph{i.e.}

\noindent
$$
\Pexc \approx (0.021,\dotsc,0.021)\,,\quad \xi = 0.4\,,
\quad \log(\sigma) = (4.8,4.6,5.9,5.1)\,,
$$

\noindent
for a  total number of days $n= 148\,401$, which is the total number of
days in the original data.   
The four-variate dependence structure is chosen as a Dirichlet mixture
distribution $h_{\dmpar}$, where

% $$
% \dmpar :\quad \Wei = (0.3, 0.3, 0.4)\,, \;
% \Mu = \begin{pmatrix}
% 0.5 & 0.2& 0.1 \\
% 0.3 & 0.1 &  0.325 \\
% 0.1 &  0.4 & 0.25 \\
% 0.1 &0.3 &  0.325 \\
% \end{pmatrix}\,, \;
% \Nu =( 30, 11, 20)\,.
% $$
$$
\dmpar :\quad \Wei = (0.25, 0.25, 0.5)\,, \;
\Mu = \begin{pmatrix}
0.1 & 0.7 & 0.1 \\
0.1 & 0.1 & 0.4 \\
0.1 & 0.1 & 0.4 \\
0.7 & 0.1 & 0.1 \\
\end{pmatrix}\,, \;
\Nu =(70, 50, 80)\,.
$$

\noindent
A radial threshold for simulation on the Fréchet scale   and the number of 
points simulated above  the latter are respectively set to 
 $r_s = -1/\log(1-\max(\Pexc))$, $n_{\text{rad.exc}} = n * 4/r_s$. 
 The remaining $n-n_{\text{rad.exc}}$ points are  arbitrarily 
scattered   below
 the radial threshold, so
 that the proportion of  radial excesses   is $4/r_{s}$, the
 exponent measure of the region $\{\mb x \in \mathbb{R}^4:
 \|x\|_1>r_{s}\}$.

Afterwards, the data set is censored following the  real data's
censoring pattern, \emph{i.e.} censoring occurs on the same
days and on the same locations  (here, a location is a coordinate
$j\in\{1,\dotsc,4\}$) as for the real data, with same
censoring bounds (on the Fréchet scale), so that the data are  observed
only if the censoring threshold is exceeded. 
Finally, in order to account for the loss of information resulting from
time dependence within the real data (whereas the
simulated data are time independent), only $n_{\text{eff}}$ data out
of $n$ are kept for inference, where $n_{\text{eff}} = \lfloor n / \text{mean
  cluster size}\rfloor = 118\,911 $ (see Section~\ref{conclusion} for
an explanation about clusters). 
The vast majority of data points (real as well
as simulated) are either
censored or below the multivariate threshold:  in the real data set,  
only $125$ multivariate excesses above threshold are recorded, among
which only $3$ have
all their coordinates of type $1$ (exact data). 
In such a context, a simplified inferential framework in which
censored data would be discarded is not an option: only $3$ points would be available for
inference.
For simulated data, the threshold 
   is arbitrarily set to the same
  value as  the one defined for  real data, \emph{i.e.} $\bthres =
  (300, 320,520,
380)$.% ,  yielding approximately
  % equal probabilities of an excess in each direction.  

The MCMC sampler described in Section~\ref{sec:MCMCalgo} and
Appendix~\ref{sec:MCMC algo} is run for each simulated
data set, yielding a parameter sample,  which approaches  the posterior distribution given the censored
data. 
To allow comparison with the default space-independent framework in
terms of marginal estimates, such as probabilities of a marginal excess,
Bayesian inference is also 
 performed in the independent model defined as follows: the marginal models
are the same as in  \eqref{marginal model}, the $Y_{j,t}, 1\le j\le
4$ are assumed to be independent while  the shape parameters $\xi_j,
1\le j\le 4$
 are, again, equal to each other. A MCMC sampler  is
 straightforwardly implemented for this independent model, following the same pattern as defined in the
 \emph{marginal moves}  for the 
 full  model (see Appendix~\ref{sec:MCMC algo}).  In the sequel,
 the full Poisson model with Dirichlet mixture dependence structure is
 referred to as the  \emph{DM Poisson
model}, or the \emph{dependent model}, as opposed to the independent
model.

\subsection{Illustration of the method with one simulated data set}\label{sec:OneSimDataset}

\begin{figure}[h]
  \centering
  \makebox{\includegraphics[scale=0.25]{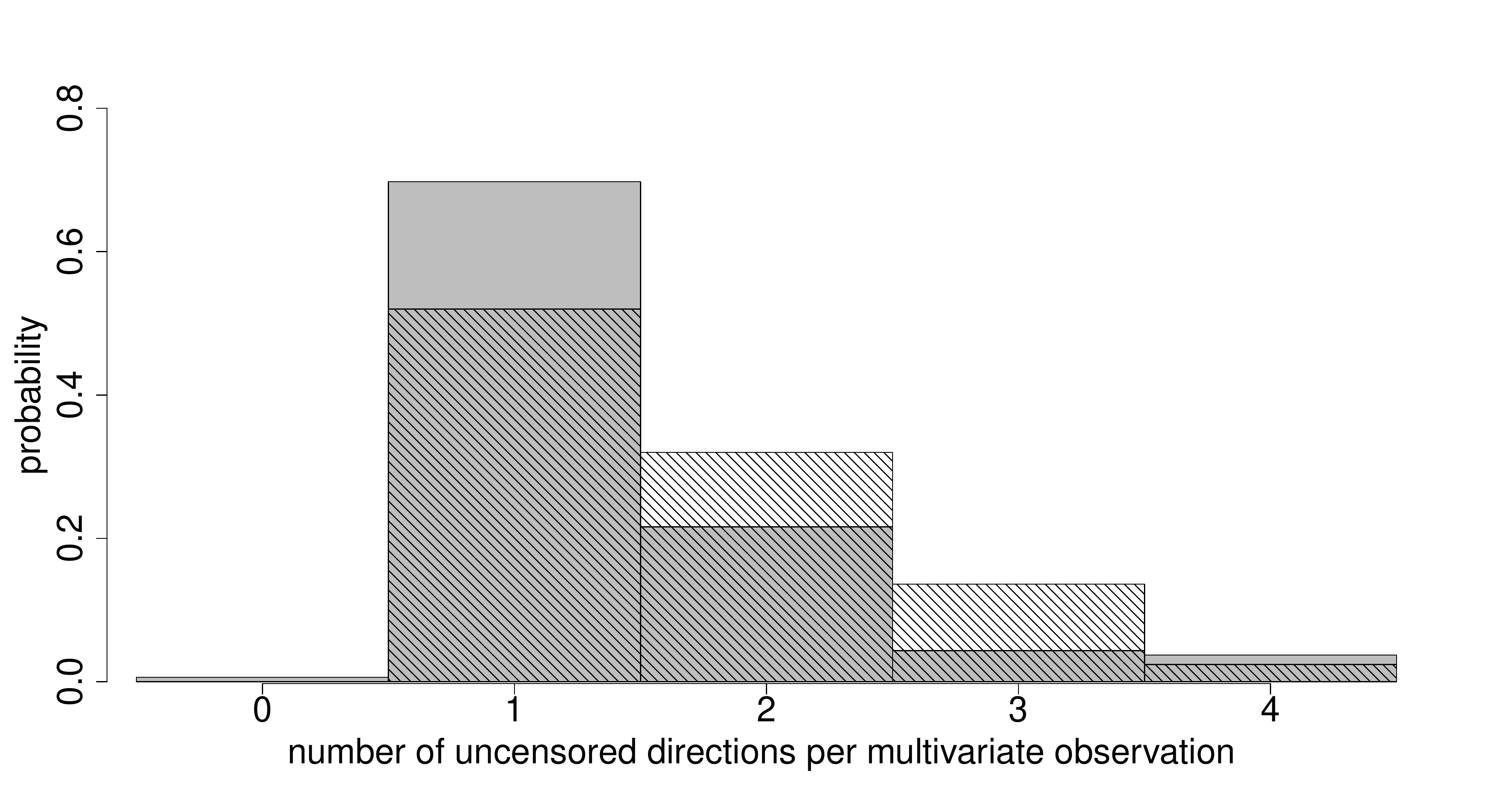}}
  \caption{  \label{fig:histTypes} Average number of exactly observed coordinates
    above threshold in each recorded multivariate excess % the simulated  data set
    % ($\#\{j:\datatype_{t_i,j} = 1\}$)
    in the real data set (hatched
    bars) and in a simulated data set (Gray bars).}
\end{figure}

% It should be noted that 
%   Only $n_\bthres =  162$ points
%  are 
% above threshold $\bthres$, from which 
The censoring pattern described above yields  a censored data set which
 resembles the real data set is term of average number of exact
 (uncensored) coordinates in each observation, as shown in  Figure
 \ref{fig:histTypes}: %  shows the allocation of the number of
% exact marginal coordinates after censoring below threshold:
most of
the extracted data have only one exact coordinate. 

\begin{figure}[h]
  \centering
  \makebox{\includegraphics[scale=0.38]{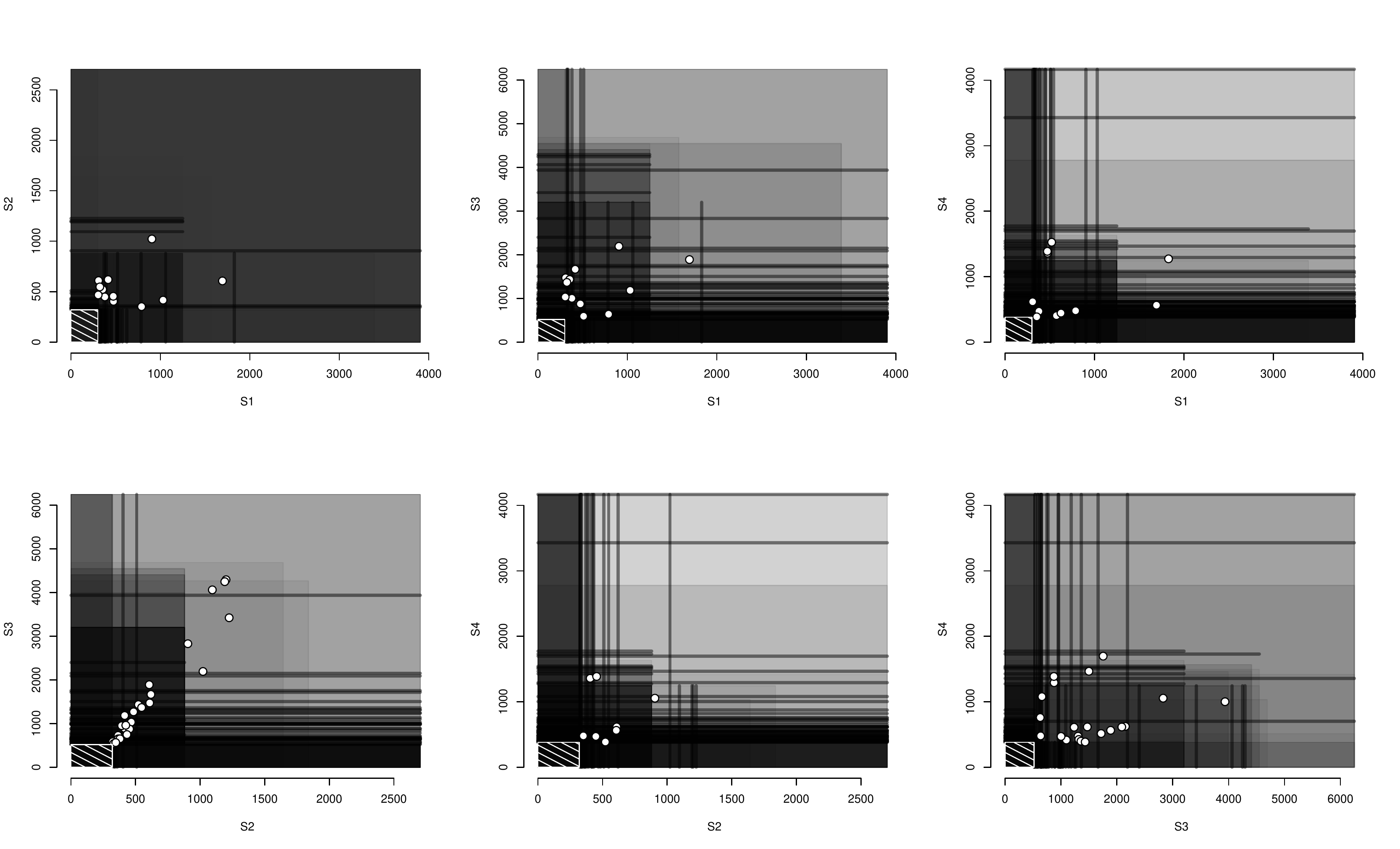}}
  \caption{\label{fig:bivarCensdata}
    Bi-variate projection of the simulated data set after
    censoring.
    White  points correspond to  pairs for which  both coordinates are
    observed. Superimposed Gray  rectangles (resp. segments)
    represent pairs for which both coordinates (resp. one coordinate) 
    are (is) censored. The white striped rectangle is the region below the
    multivariate threshold $\bthres$. }
  
\end{figure}

In addition  to data above threshold, 
the number of threshold-overlapping blocks (made of data which position with respect to
the threshold is undetermined, see  Section~\ref{sec:censoredModel})  is 
$\nblock= 39$ in this simulated data set, with block sizes varying between $1$ and $39\,845$, 
and a  total number of `undetermined' days $n'_{\bthres} = 112\,676$. 
To fix ideas, the bi-variate projections onto the six pairs $(i<j)$ of a simulated  data set are displayed in
Figure~\ref{fig:bivarCensdata}. 

%\subsection{Inference and results}
Let us turn to results obtained with  this particular  data set. %  on the augmented
% space.
A `flat' prior was specified for the DM Poisson model parameters and
MCMC  proposals  for the DM parameter were
 set  in a similar way as in \cite{sabourinNaveau2012}, which
resulted in satisfactory convergence diagnostics after $10^6$
iterations,  see Appendix~\ref{sec:MCMCsimstudy} for details. 
Figure \ref{fig:angularPred} displays the posterior predictive for
bi-variate marginalization's of the angular density, obtained
\emph{via} equation~\eqref{marginalized_param}, together with the true density
and posterior credible sets around the estimates.  The estimated
density captures reasonably well the features of the true one and the
posterior quantiles are rather concentrated around the true density. This
is all the more satisfying that, at first view
(Figure~\ref{fig:bivarCensdata}), the censored data  used for
inference  seem to convey little information about the distribution
of the angular components.

\begin{figure}[h]
  \centering
  \makebox{\includegraphics[scale=0.35]{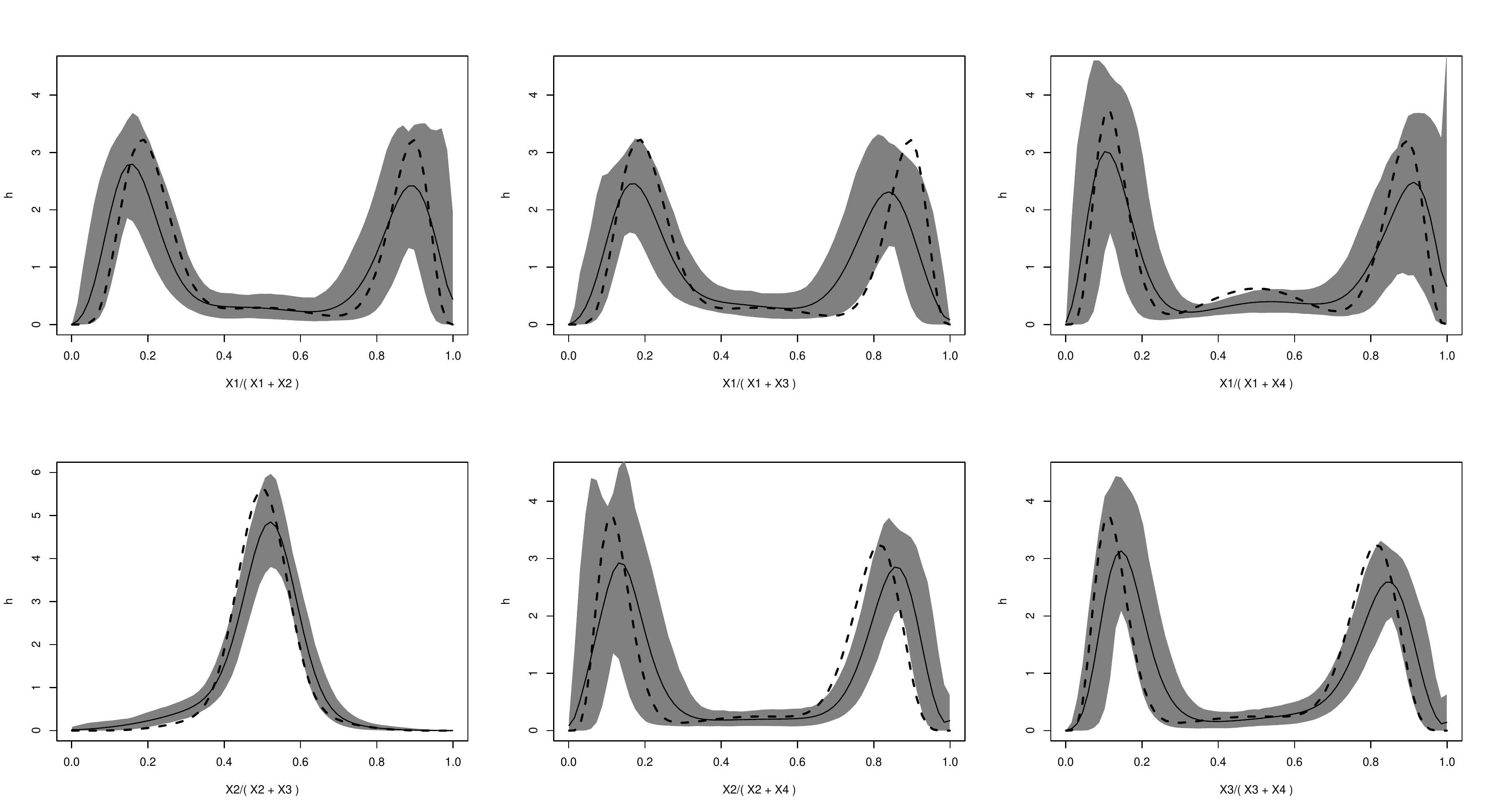}}
  \caption{  \label{fig:angularPred} Bi-variate angular predictive
    densities (thick black line)
 with posterior credible sets (Gray regions) corresponding to
 the posterior $0.05-0.95$ quantiles, together with the true angular density (dashed line).}
\end{figure}
 
\noindent 

\begin{figure}[h]
  \centering
  \includegraphics[scale=0.42]{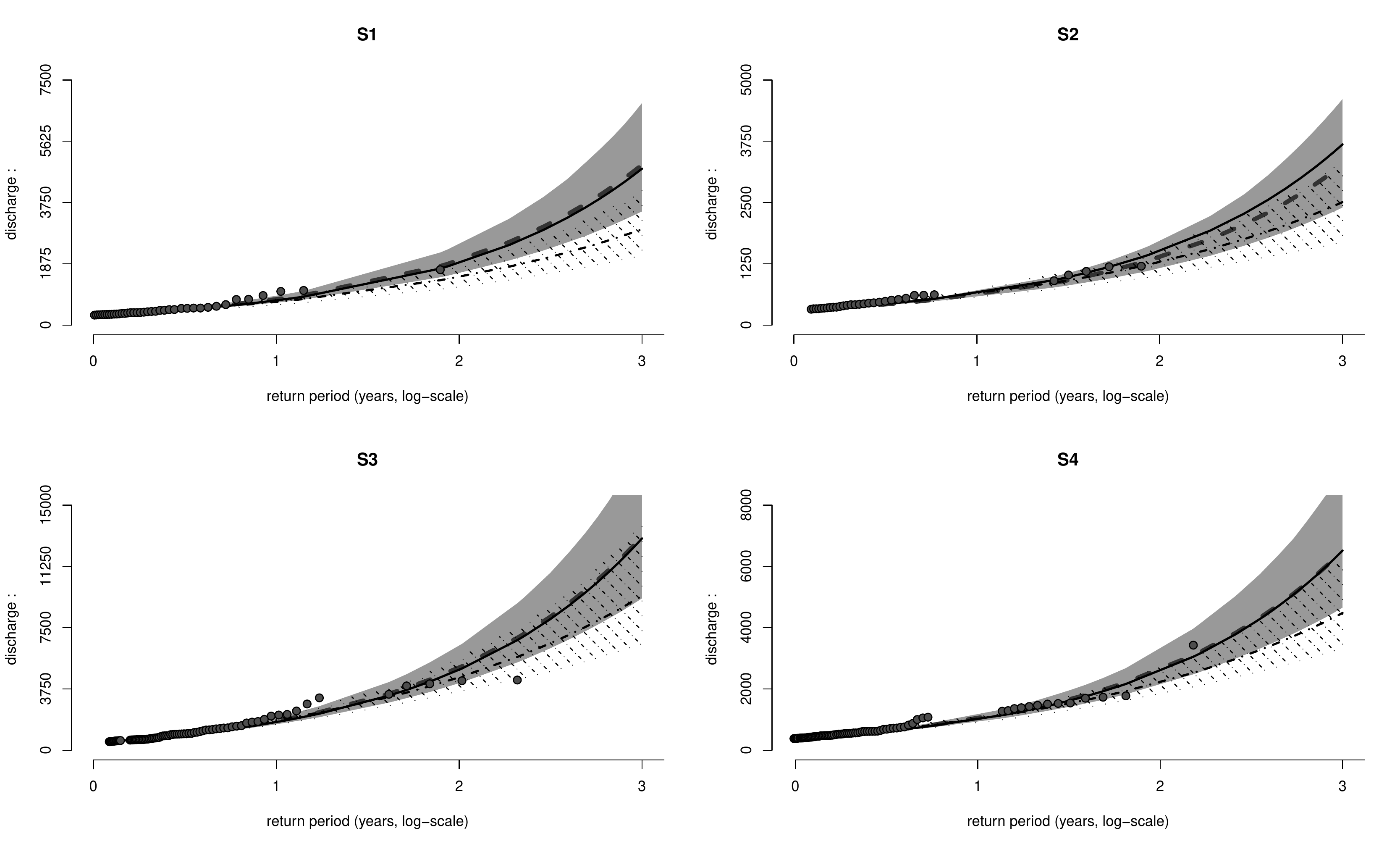}
  \caption{\label{fig:returnLevels}
    Return level plots: Quantile versus logarithm (base $10$) of the
    return period at the four locations.
    Grey points: empirical return level of observed
    threshold excesses (simulated data); Black line: true curves;
    Gray dashed line and
    shaded area: posterior mean and $0.05-0.95$-quantiles in the
    dependent model with Dirichlet mixture angular structure; 
    Black dash-dotted line and area: \emph{idem} in the independent
    model. 
  }
\end{figure}

In risk analysis,  especially in  hydrology, return level plots  (\emph{i.e.} quantile plots) are used to summarize the 
 marginal behavior of extremes. The return level $Q$ for a return period $T$
at location $j$, when marginal  data are distributed according to
$F_j^{\margpar}$ and where there is no  temporal dependence,  is usually 
defined as the $1-1/T$-quantile of 
$F_j^{\margpar}$. Figure~\ref{fig:returnLevels} compares the  return
levels obtained both in
the  dependent and the independent models, together with the true
 return levels. %  for the simulated data set. 
The posterior estimates in the dependent model are very close to the
truth, relatively to to the size of the credible intervals. In
contrast, estimation in the independent model under-estimates the
return levels, and the true curve lies outside the posterior quantiles
at two locations out of four. This single example  is however  not enough to conclude
that the dependence structure improves significantly marginal
estimation. The absence  of
significant improvement (or deterioration) of marginal estimates  is
indeed  one of the conclusions of
 the next subsection.

\subsection{Simulation study with 50 data sets}
\label{sec:sim50}

The aim of this section is to %  investigate the question of a possible improvement of marginal
% estimates  brought by  specifying a  dependence structure model
% (\emph{i.e.} a Dirichlet mixture angular measure), and to
verify
that the  posterior distribution of the DM Poisson model parameters
 is reasonably informative,  even with censored  data.  
The procedure described above is applied to generate independently  50 data sets.%  generated
% independently. 

\paragraph{\textbf{Marginal predictive performance} }
The model's ability to  estimate of the probability of a marginal
excess is first investigated. Large thresholds $(V_1,\dotsc,V_4)$ are
specified so that their 
true marginal probability of exceedance is  $P_0 = 1/(10*365)$, an approximate ten
years return level. % In the framework of scoring rules  \citep{gneiting2007strictly} 
The quantities of interest are   the posterior distributions of $\Delta_j(\theta)=
\PP_\theta(Y_j>V_j)$, 
which we hope to  be concentrated around $P_0$. 
Each posterior samples $(\theta_\iota)_{\iota\in 1 ..N}$ issued by a MCMC
algorithm is transformed into a series  of exceedance probabilities, 
$  (\delta_{j}(\theta_\iota))_{\iota\in 1..N}$ which empirical
distribution $\hat F_j = \frac{1}{N}\sum_{\iota=1}^N 
\mathds{1}_{\delta_{j}(\theta_\iota)}(\point) $  approximates  the  posterior
distribution of $\Delta_j$. %    (which true
The performance of the posterior may then be investigated in terms of
posterior quadratic loss, 

\noindent
 $$
\begin{aligned}
  \text{QL}(\hat F_j) &=  \big(\PE_{\hat F_j} (\Delta_j) - P_0\big)^2
+  
\PV_{\hat F_j}(\Delta_j)\,.%  \\
\end{aligned}
$$

\noindent 
This loss  %  standard
% scoring rules  can be useed to compare the performance of the
% probabilistic forecasts $\hat F_j$'s in the dependent
% and independent models, namely, 
corresponds to the 
the predictive model choice
criterion (PMCC) \citep{laud1995predictive}. In the framework of scoring rules
\citep{gneiting2007strictly} the PMCC is not a `proper score'  in
  general, but it is so when the true probability distribution of the
  quantity of interest  $\Delta_j$ is a Dirac mass,
  which is the case here (Dirac mass at $P_0$).  %  and the interval score (IS$_\alpha$) at level 
 % $\alpha = 0.1$, 
 % where $q_{\alpha,l}, q_{\alpha_u}$ are the $\alpha/2$ and $1-\alpha/2$
% quantiles of the forecast $\hat F_j$.  

After $1.10^6$ MCMC iterations,
at least one chain (out of six chains run in parallel for each
simulated data set) passed the Heidelberger and Welch's stationarity
tests \citep{heidelberger1983simulation} at level $10^{-4}$, for each
data set.
% The PMCC and IS scores of the dependent and
% independent models for these datasets were
% computed. 
Table~\ref{tab:QLdiff} gathers, for the  margins
 $j\in\{1,\dotsc,4\}$,  the mean and standard
deviation of the QL 
scores, normalized by the (squared) true probability $P_0^2$ for readability,

$$
\begin{aligned}
\overline{\text{QL}}_j & =  \frac{\text{QL}(\hat F_{j})}{P_0^2}\,. % \\  
% \overline{\text{IS}}_j & =  \frac{\text{IS}_\alpha(\hat F_{j,\text{dependent}}) -
% \text{IS}_\alpha(\hat F_{j,\text{dependent}})}{P_0}\,, \\  
\end{aligned}
$$ 

\begin{table}[h]
%$\PP(Y_j\ge V_j)$
\begin{tabular}{c|cccc}
$\overline{\text{QL}}_j$  & 1 & 2 & 3 & 4 \\ 
\hline
 mean  & 0.55 & 0.17 & 0.22 & 0.32 \\ 
  standard error & 0.64 & 0.14 & 0.18 & 0.30 \\ 
  first quartile & 0.14 & 0.08 & 0.08 & 0.10 \\ 
   third quartile  & 0.77 & 0.20 & 0.32 & 0.45 \\ 
  \end{tabular}\vspace{0.5cm}
  \caption{Normalized  scores  $\overline{\text{QL}}_j$ for the
    marginal probability of an excess: mean and
    standard deviation, first and third quartiles over the  $50$
    data sets. Column $j$ corresponds to an excess at location $j$
    ($1\le j \le 4$).  }
  \label{tab:QLdiff}
\end{table}

\noindent
Although the variability of the scores   (standard deviations)  is
relatively high, normalized  third
quartile less than one 
indicate that the posterior distribution concentrates in regions where
the  probability of a marginal excess  is  of the same order of
magnitude as the true probability. 

\begin{figure}[h]
  \centering
  \includegraphics[scale=0.4]{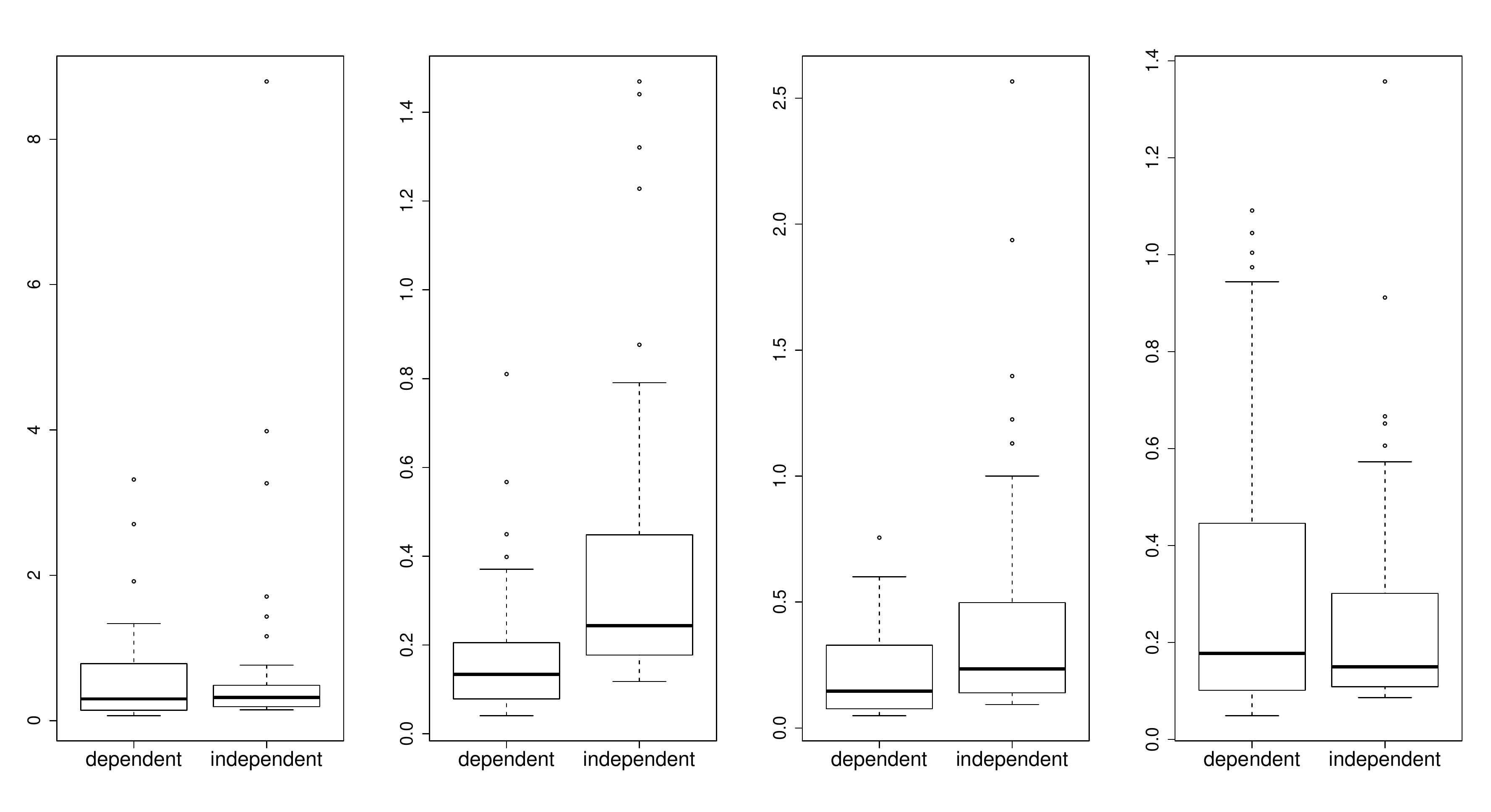}
  \caption{Normalized QR scores in the dependent Dirichlet-Poisson
    model  versus the independent model, at the four different
    locations, for 50 simulated datasets}
  \label{fig:QRboxplots}
\end{figure}

In
order to 
verify  that introducing a rather complex dependence structure model
does not deteriorate the marginal estimates, the same quadratic loss
score is computed with samples issued from the independent
model. In view of Figure~\ref{fig:QRboxplots}, displaying  box-plots of the scores
computed in both models (dependent \emph{versus} independent), there
is no significant difference  between the two models in terms of
marginal estimation. 

\paragraph{\textbf{Estimation of the angular measure}}
To assess the performance of the DM Poisson model in terms of
estimation of the dependence structure of extremes, a similar
scoring procedure as above is followed,  with  quantities
of interest $\Delta_j$ defined as the probability of a \emph{joint} excess
of the $V_i$'s, given a marginal excess of $V_j$, 

\noindent
$$
\tilde\Delta_j = \PP(Y_1 \ge V_1,\dotsc Y_4\ge V_4\; | Y_j \ge V_j)\,.
$$

\noindent
These quantities do not have an explicit expression in the DM 
 model, but are easily  approached by standard Monte-Carlo sampling. 
Lacking a reference model in this context (the independent one  is
obviously unable to predict these quantities), only  the  scores in the
dependent model are available. % However, the  QL 
% provides an order of magnitude of the accuracy and precision of the
% estimation, as it is simply the quadratic risk of the posterior
% distribution (sum of  biais and a variance term).
Table~\ref{tab:QL_dep} summarizes  the results in terms
of  normalized scores
$$\tilde{\text{QL}}_j =  \frac{
\text{QL}(\tilde \Delta_j )}{P_j^2}\,, $$ 
where $P_j$ is the true conditional probability of a joint excess. 

\begin{table}[h]
  \centering
%$\PP\big(Y_1\ge V_1,\dotsc, Y_4\ge V_4 \; |\; Y_j\ge  V_j\big)$
%$\PP\big(Y_1\ge V_1,\dotsc, Y_4\ge V_4 \; |\; Y_j\ge  V_j\big)$ 
\begin{tabular}{c|cccc}
$\tilde{\text{QL}}_j $
 & 1 & 2 & 3 & 4 \\ 
\hline
 mean & 0.06 & 0.18 & 0.25 & 0.05 \\ 
  standard error & 0.05 & 0.11 & 0.13 & 0.05 \\ 
  first quartile & 0.02 & 0.11 & 0.17 & 0.01 \\ 
  third quartile & 0.10 & 0.23 & 0.33 & 0.06 \\ 
  \end{tabular}
\ \\

\vspace{0.5cm}
  \caption{Normalized QL  scores $\tilde{\text{QL}}_j$  for conditional  probabilities  of a
    joint 
    excess: mean, standard deviation, first and third quartiles over
    the 50 simulated datasets. Column $j$ corresponds to conditioning
    upon an excess at  location $j$.}
    \label{tab:QL_dep}
\end{table}
A  comforting  fact is that the dependence structure seems to be even  better
estimated than the marginal distributions of extremes, in view of
Tables~\ref{tab:QLdiff} and \ref{tab:QL_dep}.

\subsection{Real data}\label{Gardons}
\begin{figure}[h]
  \centering
  \includegraphics[scale=0.3]{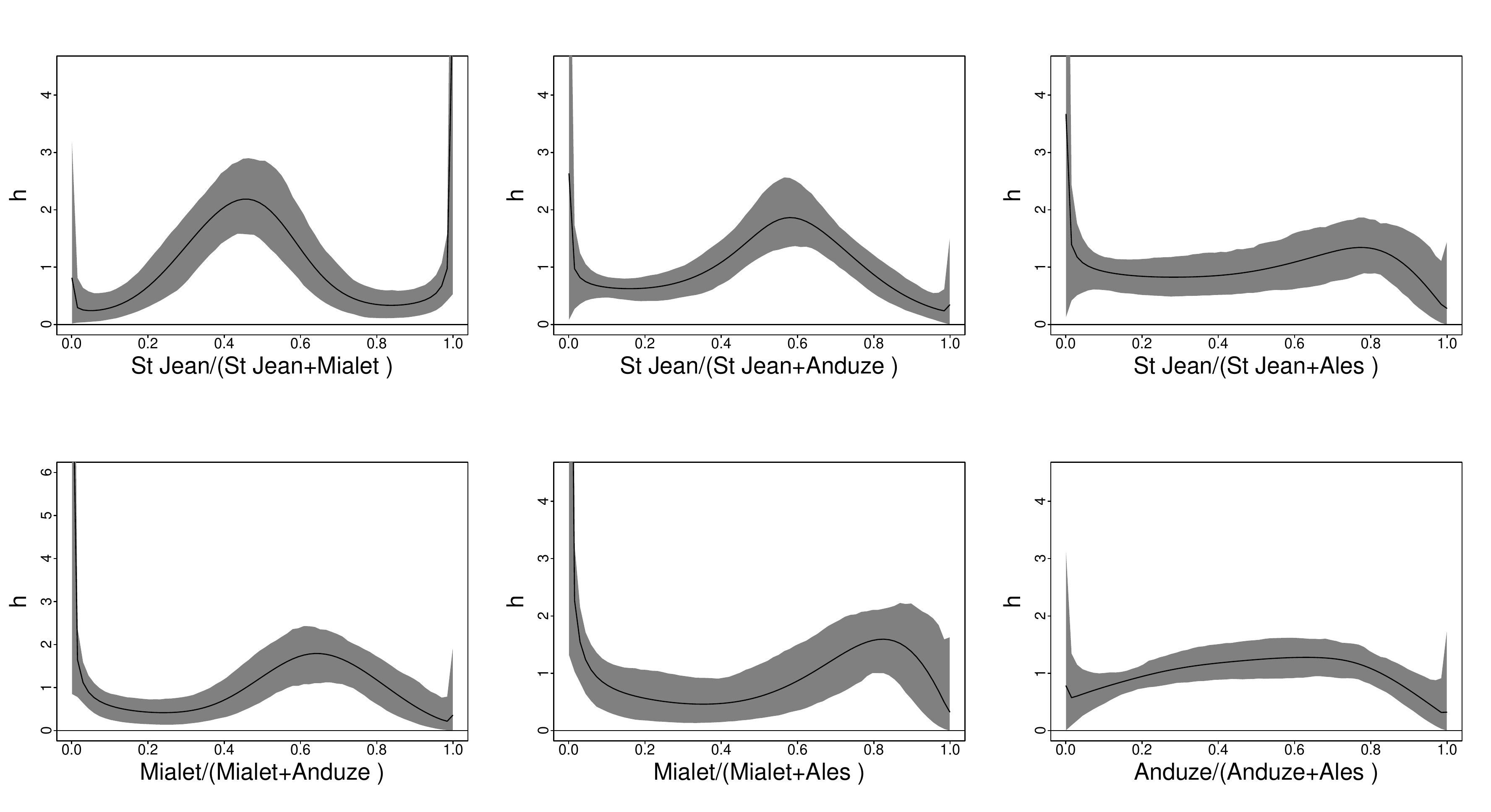}
  \caption{Bi-variate angular measure for the hydrological data set:
    posterior predictive density (black line) and posterior $0.05$ and
  $0.95$ quantiles.}
  \label{fig:angpredGardons}
\end{figure}

% This section briefly sketches some results obtain with the
%.
 Analyzing the  hydrological data set presented  in the introduction
 requires an additional declustering step: a multivariate
 run-declustering scheme was implemented. A cluster starts when one
 component exeeds the threshold. It and ends when, during $\tau$
 consecutive days, all components are below there respective
 threshold, or (when censoring is present) with undertermined
 position, so that the observer can not ascertain that an excess
 occurred. The lag parameter $\tau=3$ was set by considering
 stability regions  of
 the estimates, and additional physical characteristics of the
 hydrological catchment, see \cite{sabourin:hal-01087687} for details. After-wise the model is
 fitted to  the extracted four-variate 
cluster maxima. Again, $6$ parallel MCMC's of length $10^6$ are run,
with satisfactory convergence diagnostic after a $2\, 10^5$ burn-in
period. The posterior mean estimates of marginal parameters are close
to the maximum likelihood estimates mentioned in
subsection~\ref{sec:expeSetting}.  
Figure~\ref{fig:angpredGardons} displays bi-variate versions
of the  four-variate posterior predictive angular measure, together with point-wise
$90\%$ posterior credibility sets. 

The inferred dependence structure is rather complex, which fosters the
use of such a flexible semi-parametric model. 
The posterior credible bounds of the angular density are relatively
narrow around the posterior predictive in the central regions of the
simplex (which is the segment $[0,1]$ for bi-variate data) and  indicate
that asymptotic dependence is present. On the other hand, high levels (even unboundedness) of the predictive
density near certain edges  indicates a `weakly asymptotically dependent
regime', for the considered pair, \ie\ a regime where one component
may be large while the other one is not. Possibly, some `true'
asymptotic angular mass is concentrated on these edges, which translates
in the Dirichlet mixture model (which only allows mass on the
topological interior of the simplex)  into high densities near the
edges.  The observed
widening of posterior  credible regions near the edges where the
density is unbounded is not surprising : an unbounded density
corresponds to Dirichlet parameters with components $\nu\mu_i<1$, for
which a small variation of the parameter value induces a large variation
of the density near the edges. 

%\clearpag

\section{Conclusion}\label{conclusion}
In this work, a flexible 
semi-parametric Bayesian inferential scheme is 
implemented  to estimate
the joint distribution of excesses above multivariate high 
thresholds,  when the data are  censored. A 
simulation example  is designed on the same pattern as  a real
case borrowed from hydrology. 
Although the
tuning of the \MCMC\ algorithm requires some care, 
taking into account all kinds of observations for various censoring
bounds 
allows to obtain satisfactory  estimates, despite  the loss of  
information relative to the angular structure  induced by
the censoring process. In particular, accurate enough estimations
of quantities of interest such as marginal or conditional
probabilities of an excess of a large threshold  can be
obtained.  \\
The main methodological novelty consists in taking  advantage of the
conditioning and marginalizing properties of the Dirichlet
distributions, in order to simulate augmentation data which
`replace' the missing ones. 
Also, exponential terms in the likelihood with no explicit expressions
are handled  by
sampling well  chosen functionals of  augmentation Poisson
processes. This new inferential framework 
opens the road to statistical analysis  of the extremes of data
sets that would  otherwise  have been deemed unworkable.

%\footnote{ 
% In the present paper,  temporal dependence is not treated. In practice, time-dependent
% series  may still be analysed
% using  `declustering' methods, which  allow to fit the model to
% cluster maxima instead of the raw data. This classical approach assumes
% that only short-term dependence is present, \emph{i.e.} that a condition of  mixing type holds 
% \citep[condition
% D, see \emph{e.g.}][]{leadbetter1983extremes}, so that  cluster maxima
% may be treated as independent observations. In particular, this is
% the approach adopted in a forthcoming paper analysing data from the
% Gardons (France). 
%}, 
% Here, the Poisson Laplace transform is key to the choice of
%the functional. 

\section*{Acknowledgments}
Part of this work has been supported by the EU-FP7 ACQWA Project \\
(\texttt{www.acqwa.ch}), by the PEPER-GIS project, by the ANR-MOPERA project,
by the ANR-McSim project and by the MIRACCLE-GICC project. 
The author would like to thank Benjamin Renard for providing the
hydrological data that motivated this work and for his useful
comments,   and  Anne-Laure Fougères  and Philippe Naveau for their
advice and interesting discussions we had during the writing of this
paper.

\appendix

\section{Poisson likelihood of uncensored data}
\label{ap:simpleLikelihood}

Consider a      Dirichlet mixture density  $h=h_{\dmpar}$ as in
\eqref{dirimixDens}.  The density of the exponent measure  in Cartesian
coordinates is, using \eqref{eq:cartesianExp} and the expression of
the Dirichlet density \eqref{diriDensity},

\noindent
\begin{equation}\label{intensityDM}
\frac{\ud\expmeas_{\dmpar}}{\ud \mb x } (\mb x)=
\dimens\sum_{m=1}^k\left\{
\frac{\wei_m\Gamma(\nu_m)}{\prod_{j=1}^{\dimens}
  \Gamma(\nu_m\,\mu_{j,m})}
\prod_{j=1}^{\dimens} x_j^{\nu_m\mu_{j,m}-1} 
\left(\sum_{j=1}^{\dimens} x_j \right)^{-(\nu_m+1)}  \right\}\,.
\end{equation}

\noindent
% In the simplified context of Section~\ref{sec:model} (no censoring,
% marginal \cdf known below
% threshold), % case where the $\mb Y_{j,t}$'
%% s are all  observed
%   and where  the 
% marginals $F_j$'s below threshold are known, 
% the likelihood  of the
% Poisson process % formed by transforming and normalizing  the $\mb Y_t$'s
% observed on   $A_{\bthres} $
% $= [0,\infty]^d \setminus [0,\thres_1]\times\dotsb\times[0,\thres_%d]$ 
% is 
% \begin{equation}\label{eq:fullLikelihood}
% \begin{aligned}
%   \mathcal{L}_{\bthres} \left( \{\mb y_t\}_{1\le t
% \le n}, \theta
%   \right) 
%   & \propto e^{-n\,\expmeas_{\dmpar}(A_{\bfthres})}
%   \prod_{i=1}^{n_{\bthres}} \Big\{
%   \frac{\ud\expmeas_{\dmpar}}{\ud \mb x} ( \mb x_{t_i})%T^{\margpar}(\mb Y_{t_i}))
% \prod_{ j: y_{j,t_i} > \thres_j} J_j^{\margpar}(y_{j,t_i})
%   \Big\}\,,
% \end{aligned}
% \end{equation}
% where  $\mb X_t = \mb T^\margpar(\mb Y_t)$ 
% and $ \frac{\ud\expmeas_{\dmpar}}{\ud\mb x}$ is as in  \eqref{intensityDM}.
The likelihood expression \eqref{eq:fullLikelihood} is simply that of
a  % then obtained
% as a classical
Poisson process  on the region $[0,1]\times A_{\bfthres,n}$. 
Recall that, for a Poisson process with
intensity $\mu$ on a region $A$, the likelihood of $n$ points $(Z_1,\dotsc,Z_n)$
observed in  $A$ is proportional to $e^{-\mu(A)} \prod_{i=1}^n\frac{\ud \mu
}{\ud z }(Z_i)$. 

 The exponential term
$e^{-n\,\expmeas_{\dmpar}(A_{\bfthres})}$ in \eqref{eq:fullLikelihood}
follows from the homogeneity property of $\lambda_\dmpar$, 
 $$
\begin{aligned}
e^{- %(\int_{0}^1 \uds)
\leb([0,1])\,.\,
\expmeas_{\dmpar}(A_{\bfthres,n})} = e^{-n \expmeas_{\dmpar}(A_{\bfthres})}\,.
\end{aligned}
$$

The terms $J_j^\margpar$ are the inverse Jacobians  of the marginal  transformations 
$T_j^\margpar :  Y_j \mapsto X_j$, \ie 

\noindent
\begin{equation*}\label{jacobian}
 J_j^{\margpar}(y_j) =% \frac{1}{n}
\sigma_j^{-1}(\pexc_j )^{-\xi_j}
x_{j}^2 e^{\frac{1}{x_{j}}} \left[1-e^{\frac{-1}{x_{j}}}
\right]^{1+\xi_j} \,,
\end{equation*}
 where $x_j = \mathcal{T}_j^{\margpar}(y_j)$,

\section{Integration of the exponent measure along directions of missing components}
\label{sec:integrMissing}
The  exact expression for  the integral of
the exponent measure's density   in \eqref{censoredLkl_full}  along the axes $[0,\infty]$ corresponding to missing
coordinates  is given below.  
% The dimension of integration in \eqref{censoredLkl_full} for data in the
% failure region can be reduced as far as `missing coordinates' are
% concerned, which corresponds to marginal integration bounds equal to
% $(0,\infty)$.
% , which corresponds to
% `missing coordinates'. 
% been reduced with  expression
% \eqref{intensityDM_missing}
% for 
% partial integration in the directions of missing components ($\tilde \datatype_{j,t_i}
% = 0$). 
% The case where some coordinates are missing (not observed at all) 
% is easily covered:
Let 
$\mathscr{D} = \{j_1,\dotsc,j_r\}$ be  the non missing coordinates
($r<\dimens$).    Integrating $\frac{\ud\expmeas_{\dmpar}}{\ud\mb x}$ over $(0,\infty)$ in
the missing directions 
$\mathscr{D}_0 = i_1,\dotsc,i_{d-r}$ 
yields the marginal 
density of $\expmeas$  
% Then, the exponent measure \eqref{intensityDM} has a density
with respect
to the Lebesgue measure on the vector space spanned by
$\mathscr{D}$, \ie 
$\frac{\partial\expmeas_{\dmpar}  (\mb x)}{\partial x_{j_1}\dotsb\,\partial
  x_{j_r} } $. 
% which is obtained by 
% % \eqref{intensityDM}  in  the missing directions
% $\mathscr{D}_0 =\{s_{1},\dotsc,s_{\dimens-r}\} =\{1,\dotsc,d \}\setminus\mathscr{D}
% $  
% with integration bounds set to $(0,\infty)$
With a DM angular measure,  the integral has an
analytic expression, which is, using  \eqref{diriDensity} and
\eqref{dirimixDens},  

\begin{equation}\label{intensityDM_missing}
  \begin{aligned}
    \frac{\partial\expmeas_{\dmpar}  (\mb x)}{\partial x_{j_1}\dotsb\partial
      x_{j_r} } &= 
    \int_{\{\mb z : z_j = x_j (j \in \mathscr{D}), z_i\in\mathbb{R}^+
      (i\in\mathscr{D}_0)\}} 
    \frac{\ud\expmeas_{\dmpar}}{\ud \mb x } (\mb z)  \ud z_{i_1},\dotsc,\ud z_{i_{\dimens-r}}\\
    &= r \sum_{m=1}^k 
    \left(
      \frac{\wei_m^0 %%\left(1-\sum_{s\in\mathscr{D}_0} \mu_{s,m}^0 \right)
        \Gamma(\nu_m^0)}{\prod_{j \in \mathscr{D}}
        \Gamma(\nu_m^0\mu_{j,m}^0)}
      \prod_{j\in\mathscr{D}} x_j^{\nu_m^0 \mu_{j,m}^0-1} 
      \left(\sum_{j\in \mathscr{D}} x_j \right)^{-(\nu_m^0 + 1)}  
    \right)
    \,,
  \end{aligned}
\end{equation}
with 
\begin{equation}\label{marginalized_param}
  \begin{gathered}
    \nu_m^0 = \nu_m  (1-\sum_{i\in\mathscr{D}_0}
    \mu_{i,m})\;,\quad 
    \Mu_{m}^0 =
    (1-\sum_{i\in\mathscr{D}_0} \mu_{i,m})^{-1}\Mu_{m}\;,\\
    \wei_m^0 = \frac{d}{r}(1-\sum_{i\in\mathscr{D}_0} \mu_{i,m})
    \wei_m \;.
  \end{gathered}
\end{equation}
This is the spectral measure associated with  another 
angular
DM distribution on $\mathbf{S}_r$  with parameter  $\psi^0 = (\nu^0_{1:k}, 
\Mu_{1:k}^0,  \wei_{1:k}^0)$. 
The censored likelihood \eqref{censoredLkl_full} can thus be re-written as  

\noindent
\begin{equation}\label{censoredLkl}
  \begin{aligned}
    \mathcal{L}_{\bthres}(\mb O, \theta)
    &=\exp\Big[- \ndef\,\expmeas_{\dmpar}(A_{\bfthres} ) -
    \sum_{i=1}^{\nblock} n'_{i}
    \expmeas_{\dmpar}(A'_i)\Big]%\big([\mb 0,\tilde {\mb{R}}_{t'_i}]^c \big) \Big]
    \times\dotsb\\
    &\dotsb\,\prod_{i=1}^{n_{\bthres}} \Big\{
    \int_{\mb [\mb{\tilde L}_{t_i}^0 \mb{, \tilde
        R}_{t_i}^0 \mb ]}
    % {\mathscr{C}_{t_i}^\margpar(\mb x) \in \mb
    % {C}_{t_i}^{\margpar}}
    % \left[\mathscr{C}_{t_i}^\margpar \right]^{-1}( \mb {C}_{t_i}^{\margpar})}
    % {\mathscr{E}_{t_i}^\margpar(\mb C_{t_i}^{\margpar}) }
    \frac{\partial^{r(i)}\expmeas_{\dmpar}}{\partial x_{j_1(t_i)}\dotsb\partial
      x_{j_{r(i)}(t_i)} }(\mb x)
    \ud\genemeas_i^0 
    \prod_{ j: Y_{j,t_i} > \thres_j} J_j^{\margpar}(y_{j,t_i}) \Big\}\,,
  \end{aligned}
\end{equation}

\noindent
where  
integration is performed in the censored, non-missing directions, and
where 
  $\genemeas_i^0$ is the Lebesgue measure on the corresponding
subspace of $\mathbb{R}^d$ and $\mb [\mb{\tilde L}_{t_i}^0 \mb{, \tilde
        R}_{t_i}^0 \mb ]$ are the original bounds 
$\mb [\mb{\tilde L}_{t_i} \mb{, \tilde
        R}_{t_i} \mb ]$, modulo  the subspace of missing
      components. 
% respect to the  space spanned by $\mathscr{D}_c(i)$.

% where, for $i\le n_{\bthres}$,
% $\mathscr{E}_{t_i}^{\margpar}(\mb C_{t_i}^{\margpar}) =  \{ \mb x : \mathscr{C}_{t_i}^\margpar(\mb x) = 
% \mb {C}_{t_i}^{\margpar}\} =
% \left[\mathscr{C}_{t_i}^\margpar \right]^{-1}(
% \mb {C}_{t_i}^{\margpar}) $,

\section{Data augmentation details}
\label{sec:dataAugmentFull}
Here is detailed the construction of   augmentation data
$\mb Z = (\mb Z_{\text{above}} ,\mb Z')$, first introduced in
Section~\ref{sec:DataAugmentPoissonModel}.
\subsection{Definition of augmentation  the variable $\mb Z_{\text{above}}
  = \{\mb Z_{t_i}\}_{i\le n_{\bthres}}$ }
\label{sec:dataAugmentDetails}
Consider a Fréchet-transformed, censored observation
$\mb C_{t_i}^\margpar = (C_1^\margpar,\dotsc,C_\dimens^\margpar)$,
with  

\noindent
$$
C_j^\margpar = (\tilde \datatype_{j,t_i},  X_{j,t_i}, \tilde
L_{j,t_i}, \tilde R_{j,t_i}), $$

\noindent
 as  in Section~\ref{sec:censorThres}. % \ref{sec:censoredModel}
% and Appendix~\ref{ap: formalCensFrech}.
%
% % In a first step, let us forget about the  missing components in the censored
% data above threshold, \emph{i.e.}, assume that 
%
% \noindent
% $$\mathscr{D}_{0,t_i} = \{(j,t_i):~\tilde \datatype_{j,t_i} = 0\}
% =\varnothing.$$
%
% \noindent
% For $i \le n_{\bthres}$, 
Let 
$\mathscr{D}_c(i) = (j'_1(i),\dotsc,j'_c(i))$ be the 
  censored
 coordinates in observation $\mb C^\margpar_{t_i}$. 
The latent variables
$\mb Z_{t_i} = (Z_{j'_1,t_i},\dotsc,Z_{j'_c,t_i})$ 
%($j\in\mathscr{D}_c(i), i\le n_{\bthres}$)
are defined so as  to  `replace' those coordinates: 
 % ce censored
% coordinates in $ \mb C_{t_i}^\margpar$. 
More formally, let

\noindent 
\begin{equation}\label{barXj}
\mb{ \bar{X}}_{t_i}\,|\,\dmpar\, \sim  \frac{1}{\expmeas_{\dmpar}(A_{\bfthres})}\,
\mathds{1}_{A_{\bfthres}}(\point)\,\expmeas_{\dmpar}(\point)\,,
\end{equation}

\noindent
be an  uncensored $d$-dimensional variable with Fréchet margins and
dependence structure given by $\expmeas_{\psi}$ on $A_{\bfthres}$.
% Denoting $\mathscr{D}_1(i) = \{1,\dotsc,\dimens\}\setminus 
% % \mathscr{D}(i) \setminus
% \mathscr{D}_c(i)$  the  `exactly observed'
%  coordinates  ($\tilde\datatype_{j,t_i}~=~1$), 
Then, %fr $j \in \mathscr{D}_c(i)$,
$\mb Z_{t_i}$ is defined through its joint distribution with the $\{
X_{j,t_i}\;:\; j\notin\mathscr{D}_c(i)\}$, conditionally on $\theta$,
%(a factor of the augmented likelihood),

$$
\Big(\mb Z_{t_i}, \{ X_{j,t_i}\;:\; j\notin\mathscr{D}_c(i)\}\Big)
\,\big|\, \theta\,\quad \overset{\text{distribution}}{=} \quad \mb{ \bar{X}}_{t_i} \,|\dmpar.
$$

% Then,   %fr $j \in  \mathscr{D}_c(i)$, 
%   $\mb Z_{t_i}$ is  defined through its joint distribution with
%   $\mb O_{t_i}$, conditionally on $\theta$ (a factor of the augmented likelihood), 
%
% $$
% (\mb Z_{t_i}, \mb C_{t_i}) \,|\, \theta\, \sim  \mb{ \bar{X}}_{t_i} \,|\dmpar.
% $$
Then, conditionally on the observation $\mb
  O_{t_i}$, 

\begin{equation} \label{fullconditional}
\begin{aligned}
 \big[\mb Z_{t_i} \,\vert \,\mb O_{t_i},\theta\,\big] & =
 [(\bar{X}_{j'_1,t_i},\dotsc,\bar{X}_{j'_c,t_i}) \;|\; \mb O_{t_i},\theta ]\\
&=  \Big[(\bar{X}_{j'_1,t_i},\dotsc,\bar{X}_{j'_c,t_i})
\Big|\; \mb{\bar{X}}_{t_i}\in \mb{[\tilde L}_{t_i},\mb{\tilde R}_{t_i}\mb], \;
\dmpar\; \Big]\,.
%[\mathscr{C}_{t_i}^{\margpar}]^{-1}( \mb C_{t_i}^{\margpar}),\dmpar]\,.
% &=  \Big[(\bar{X}_{j'_1,t_i},\dotsc,\bar{X}_{j'_c,t_i}) |\big\{\bar{X}_{j,t_i}\in(\tilde
% L_{j,t_i},\tilde R_{j,t_i}), j\in\mathscr{D}_c(i)\big\}\, 
% \dotsb\\
% &\qquad\qquad\qquad \dotsb, \big\{ \bar{X}_{j, t_i}
% = x_{j,t_i}, j\in \mathscr{D}_1(i)\big\}, \dmpar\Big]\,,\\
\end{aligned}
\end{equation}
% where, by convention, $\tilde R_{j,t_i} = +\infty$ if 
% $\tilde\datatype_{j,t_i} = 2$. 

The contribution of the `augmented' data point $(\mb Z_{t_i},\mb
O_{t_i})$ to the augmented likelihood \eqref{eq:augmentedLKL} is 
 % on $\mathrm{Vect}(\mathscr{D}_c) \times \mathrm{Vect}(\mathscr{D}_1)$

\noindent
\begin{equation}
  \label{eq:jointCompletedDensity}
[\mb z_{t_i},\mb O_{t_i}|\theta ]= 
\frac{\ud\expmeas_{\dmpar}}{\ud x}(\mb{\bar x}_{t_i})
%\ud\genemeas_i 
  \prod_{ j: \datatype_{j,t_i}=1} J_j^{\margpar}(y_{j,t_i}) \,,  
\end{equation}

\noindent
where

\noindent
$$
\bar{x}_{j,t_i} = \begin{cases}
\mathcal{T}_{j}^{\margpar}(y_{j,t_i}) & \text{if } 
\tilde\datatype_{j,t_i} = 1\,,\\
%j\in\mathscr{D}_1(i) \,, \\
z_{j,t_i}& \text{otherwise  .} %j\in\mathscr{D}_c(i)\,.
\end{cases}
$$

\begin{rem}
With missing components $\mathcal{D}_0(i) = \{j: \datatype_{j,t_i} =
0\} \neq\emptyset$,  integration in the direction
$\mathcal{D}_0(i)$ can be   performed analytically (see
Appendix~\ref{sec:integrMissing}), which reduces the dimension of the
augmented data. Indeed, in such a case, the corresponding $Z_{j,t_i}$'s
need not be included in $\mb Z_{t_i}$, the
uncensored variable $\mb{\bar{X}}_{t_i}$  is  defined on the quotient spaces
$\mb E/\mathcal{D}_0(i)$ and its distribution is  proportional to
the exponent measure $\expmeas_{\dmpar^0}$ defined by
equations~\eqref{intensityDM_missing} and \eqref{marginalized_param}, with density $
\frac{\partial^{r(i)}\expmeas_{\dmpar}}{\partial
  x_{j_1(t_i)}\dotsb\,\partial x_{j_{r(i)}(t_i)} }(\point)\,.$ as in
equation~\eqref{censoredLkl}, so that 

\noindent
\begin{equation}
  \label{eq:jointCompletedDensity_integrated}
[\mb z_{t_i},\mb O_{t_i}|\theta ]= 
\frac{\partial^{r(i)}\expmeas_{\dmpar}}{\partial
   x_{j_1(t_i)}\dotsb\,\partial x_{j_{r(i)}(t_i)} }
% \frac{\ud\expmeas_{\dmpar}}{\ud x}
(\mb{\bar x}_{t_i})
%\ud\genemeas_i 
  \prod_{ j: \datatype_{j,t_i}=1} J_j^{\margpar}(y_{j,t_i}) \,,  
\end{equation}

  \end{rem}

  \subsection{Full conditional distribution of augmented 
data $[\,\mb  Z_{\text{above}} \,|\,\mb O,\fullpar\,]$}\label{ap: augmentedDistr}

The full conditionals
$[ Z_{j,t_i} |\{Z_{s,t_i} \}_{s \neq j},  \mb O_{t_i},\fullpar ]$ are
functions of truncated Beta distributions that can
easily be sampled  in a
Gibbs step of the algorithm, as shown below. %  (see Appendix \ref{ap: augmentedDistr}).

% In this section,   the full conditional distribution of the
% augmented  variables $Z_{j,t_i}$ $ (\tilde \datatype_{j,t_i}
% \in\{2,3\})$ is derived, given  the other augmented components $\mb
% Z_{r,t_i}$, ($r\neq j$), the parameter $\theta$ and the observation
% $\mb O_{t_i}$. 
In the remaining of this subsection, we omit the temporal index $t_i$. 
If  $\dmpar$ is a mixture of $k$ Dirichlet distributions,
as  in 
\eqref{dirimixDens}, then, 
for any bounded, continuous function $g$ defined on $[ \tilde
L_{j},\tilde R_{j}]$, the conditional expectation of $g(Z_j)$ is,
 up to a multiplicative constant,

\noindent
\begin{equation}
\label{egcondit}
\begin{aligned}
\mathbb{E}\left[g(Z_j) \left\vert \mb O,\{ Z_{r} \}_{r \neq j},
\fullpar  \right. \right] &=  
\mathbb{E}\left[g(\bar X_j) \left\vert \bar X_j \in[\tilde L_j,\tilde R_j], \bar X_r = x_r\,( r\neq j)
,\,\dmpar\right.  \right] \\
 & = \int_{\tilde L_j}^{\tilde R_j}
  g(x_j)h_{\dmpar}\Big(\frac{\mb x }{ \sum_{r}  x_{r}}\Big) 
 \Big(\sum_{r}  x_{r}\Big)^{-(d+1)} \ud x_j  \\
&= \sum_{m=1}^{k} \wei_m 
\underbrace{\int_{\tilde L_j}^{\tilde R_j} 
g(x_j)h_{\dmpar,m}\Big(\frac{\mb x }{ \sum_{r}  x_{r}}\Big) 
 \Big(\sum_{r}  x_{r}\Big)^{-(d+1)} \ud x_j}_{I_m}\,,
\end{aligned}
\end{equation}
% where the $\tilde R_{j}, \tilde L_{j}$'s are the Fréchet-transformed,
% censored  bounds defined in  section \ref{sec:censorThres}.

\noindent
(see equation~\eqref{barXj} for the definition of $ \bar X_j$).

Each term $I_m$ ($m\le k$ for a mixture of $k$ components) is 
\begin{equation*}
  \begin{aligned}
   I_m &= \gamma_m 
\int_{\tilde L_j}^{\tilde R_j} 
 g(x_j) \prod_{r\le d} x_{r}^{\nu_m\,\mu_{r,m}-1}
\left(\sum_{r\le d} x_{r} \right)^{-(d+1)-(\sum_{r \le
    d}(\nu_m \,\mu_{r,m}-1) )} \ud x_j \\
&=\gamma_m %%\frac{\Gamma(\nu_m)}{\prod_{r = 1}^d \Gamma(\nu_m\mu_{r,m})}
\int_{\tilde L_j}^{\tilde R_j} 
g(x_j) \prod_{r\le d} x_{r}^{\nu_m\,\mu_{r,m}-1}
\left(\sum_{r\le d} x_{r} \right)^{-(\nu_m+1)} \ud x_j\\
&= \gamma_ m\,%%\frac{\Gamma(\nu_m)}{\prod_{r = 1}^d \Gamma(\nu_m\mu_{r,m})}
\rho_j
\int_{\tilde L_j}^{\tilde R_j} 
g(x_j)x_j^{\nu_m\mu_{j,m}-1} (s_j+x_j)^{-\nu_m-1} \ud x_j\,,
 \end{aligned}
\end{equation*}
where $\gamma_m = \frac{\Gamma(\nu_m)}{\prod_{r = 1}^d
  \Gamma(\nu_m\mu_{r,m})}$,  $s_j = \sum_{r\neq j} x_r$ 
and $\rho_j = \prod_{r\neq j} x_r^{\nu_m\mu_{r,m}-1}$.
Changing  variable with 
$u = x_j/(x_j+s_j)$, the integration bounds are 

$$
 R'_{j} = \frac{\tilde R_{j}}{s_{j} +\tilde R_{j}} \,,\;
 L'_{j} = \frac{\tilde L_{j}}{ s_{j} + \tilde L_{j}},
$$
and we have 

\begin{equation*}
  \begin{aligned}
    I_m &=  \gamma_m\,  %%\frac{\Gamma(\nu_m)}{\prod_{r = 1}^d \Gamma(\nu_m\mu_{r,m})}
    \rho_j\,
    \int_{L'_j}^{R'_j} g\left(\frac{s_j u}{1-u}\right)\, 
    \left(\frac{s_j
        u}{1-u}\right)^{\nu_m\mu_{j,m}-1}
    (s_j+\frac{s_j u}{1-u})^{-\nu_m-1}
      \frac{s_j }{(1-u)^2}\ud u \\
      &=  \gamma_m\, %\frac{\Gamma(\nu_m)}{\prod_{s = 1}^d \Gamma(\nu_m\mu_{s,m})}
      \rho_j\,
      s_j^{-\nu_m(1-\mu_{j,m}) -1}   
    \int_{L'_j}^{R'_j} g\left(\frac{s_j u}{1-u}\right)\,
    u^{\nu_m\mu_{j,m} -1}(1-u)^{\nu_m(1-\mu_{j,m})}\ud u \\
  \end{aligned}
\end{equation*}
One recognizes in the integrand the unnormalized density of a Beta
random variable 
$U_m~\sim~\dbeta(a_m,b_m)$, with 

\begin{equation*}
  \label{betapars}
  a_{m} = \nu_m\mu_{j,m}\,,\quad
 b_{m}= \nu_m(1-\mu_{j,m})  +1\,;
\end{equation*}
Let $\mathrm{IB}_{a,b}(x,y)$ denote the incomplete Beta function (\emph{i.e.}
the integral of the Beta density)  between truncation bounds $x$ and
$y$. The missing normalizing constant in the integrand is 

$$
\begin{aligned}
D_m &=
\frac{\Gamma(a_m+b_m)}{\Gamma(a_m)\Gamma(b_m)\mathrm{IB}_{a_m,b_m}(L'_j,
  R'_j)} \\
&=\frac{\Gamma(\nu_m)}
{\Gamma(\nu_m\mu_{j,m})\Gamma(\nu_m(1-\mu_{j,m}))}
\frac{1}{\mathrm{IB}_{a_m,b_m}(L'_j, R'_j)(1-\mu_{j,m})}
 \end{aligned}
$$

Finally, we have
$$
\begin{aligned}
  I_m =
C_m .D_m%\frac{1}{D_m} 
\int_{L'_j}^{R'_j} g(\frac{s_j u}{1-u})\,
u^{\nu_m\mu_{j,m} -1}(1-u)^{\nu_m(1-\mu_{j,m})}\ud u 
\end{aligned}
$$
with

\begin{equation}
  \label{eq:multweights}
 C_m =(1-\mu_{j,m})\frac{ \Gamma(\nu_m(1-\mu_{j,m}))}
{\prod_{s \neq j}
      \Gamma(\nu_m\mu_{s,m})} \mathrm{IB}_{a_m,b_m}(L'_j, R'_j)\,
\rho_j\,
s_j^{-\nu_m(1-\mu_{j,m}) -1} , 
\end{equation}
 so that that the conditional expectation \eqref{egcondit} is that of
 a mixture distribution, 

\begin{equation*}
  \label{eq:espeFullCondit}
  \mathbb{E}\left[ g(Z_j)|\mb O,\{Z_{r,r\neq j}\} \right] = 
\sum_{m = 1}^{j}\, p'_m \, 
\mathbb{E}\left[g( \frac{s_j U_m}{1-U_m})\right]
\end{equation*}
with
weights $(p'_m)_{m\le k}$,
\begin{equation}
  \label{eq:weightsConditmixt}
  p'_m = \wei_m\, C_m \,,
% (1-\mu_{j,m})\frac{ \Gamma(\nu_m(1-\mu_{j,m}))}
% {\prod_{s \neq j}
%       \Gamma(\nu_m\mu_{s,m})} \mathrm{IB}_{a_m,b_m}(L'_j, R'_j)\,
% \rho_j\,
% s_j^{-\nu_m(1-\mu_{j}) -1} ,
\end{equation}
 where $C_m$ is given by equation \eqref{eq:multweights}.

As a conclusion,  the conditional variable $[Z_j |\mb O,Z_{s\neq
  j},\fullpar]$ is a mixture distribution  of $k$ components 
\begin{equation}
  \label{eq:mixtureConditDistr}
\left(p'_m, V_{j,m} = \frac{s_j U_m}{1-U_m} \right)_{ 1\le m\le k}\,.
\end{equation}

In presence of missing coordinates,
\eqref{eq:mixtureConditDistr} 
still holds, up to replacing  the mixture parameters $(\Wei,\Mu, \Nu)$ 
with $(\Wei^0, \Mu^0, \Nu^0)$ as in
equation \eqref{marginalized_param}.% ,  
% with   multiplicative corrective factors
% $1-\sum_{j\in\mathcal{D}_0(i)}\mu_{j,m}$.

\subsection{Augmentation  Poisson process 
$\mb Z'= \{ \{\mb Z'_i\}_{i\le  \nblock}, \mb Z'_{\bfthres}\}$  and weight function $\varphi$}\label{sec:augmentExp}
% plays a central role in the choice of 
% $\varphi$. 

Let us define a
region $E_{\bfthres,\ndef}= \{\mb x \in (\mathbb{R}^+)^d : \|\mb x\|_1 > 
\min_j(\frac{\fthres_j}{\ndef} ) \} $, so that 
$A_{\bfthres,\ndef} \subset  E_{\bfthres,\ndef}$. 
Choose a multiplicative constant $\tau>0$ and define a Poisson
intensity measure 
$\expmeas'(\point) = \tau \,\expmeas_{\dmpar}(\point)$. 
The augmentation process $\mb Z'_{\bfthres}$ is a   Poisson processes
which is defined together with  $\varphi_{\bfthres}$ by 

\noindent
\begin{equation}
\label{defEf}
  \begin{cases}
\mb Z'_{\bfthres}\sim \PRM(\expmeas') \text{ on } E_{\bfthres,\ndef}\,,\\
\varphi_{\bfthres}(\mb Z'_{\bfthres}) =  (1-1/\tau)^{\mb Z'_{\bfthres}(A_{\bfthres,\ndef})}\,,
 \end{cases}
\end{equation}

\noindent  where $\mb Z'_{\bfthres}(A_{\bfthres,\ndef})$ is the number of  points forming
$\mb Z'_{\bfthres}$ which hit  $A_{\bfthres,\ndef}$. 
Full justification and simulation details are given in the next
subsection (Appendix~\ref{consistencyPoisson}).

  % In
% particular, it is shown that equation \eqref{defEf} implies the  consistency condition
% \eqref{eq:consistencyLikelihood}.
 Let $ \{\mb{x}'_{\bfthres,s}=(r'_{\bfthres,s}, \mb
 w'_{\bfthres,s})\}_{s\le N'_{\bfthres}} $ be the points of $\mb Z'_{\bfthres}$ in
  $E_{\bfthres,\ndef}$, the density of $\mb Z'_{\bfthres}$ over
  $E_{\bfthres,\ndef}$, which contributes  to
  the augmented likelihood \eqref{eq:augmentedLKL},  is 

\begin{equation}\label{densityZu}
 [\mb z'_{\bfthres}\mid \dmpar] = \frac{1}{
%%   \mb z'_{\bfthres}(E_{\bfthres,\ndef})!
N'_{\bfthres}!
}\; e^{\frac{-\ndef\,\tau\,\dimens  }{\min_{j\le d}
     \bfthres_j } } 
\prod_{s= 1}^{N'_{\bfthres}}
%Z'_{\bfthres}(E_{\bfthres,\ndef})} 
% \frac{\ud\expmeas_{\dmpar}}{\ud \mb x } ({\mb X}'_{\bfthres,s}) 
\frac{\tau\,\dimens}{(r'_{\bfthres,s})^2} 
h_{\dmpar}(\mb w'_{\bfthres,s})\,.
%%{Z'(A_{\bfthres,n})}\,.
\end{equation}

 The processes $\mb Z'_i$'s and the weights $\varphi'_i$'s
are defined similarly,  replacing $\ndef$ with $n'_i$ and
$\fthres_j$ with $\tilde{R}_{j,t'_i}$ ($1\le i\le \nblock$).

\subsection{Consistency of the augmentation model .}\label{consistencyPoisson}

The augmentation process  $\mb Z'$ and the weight
function $\varphi$ %  (accounting for  the exponential factors in the
% likelihood) 
have been constructed with a hint towards  using  the general expression of the   Laplace transform
 of a  Poisson process to prove consistency of the augmented posterior.  

\begin{proposition}\label{prop:consistency}

Define  the factors 

\noindent
$$[\,\mb z_{t_i}, \mb
O_{t_i}\,|\,\fullpar\,] \,(i\le n_{\bthres})\quad  \text{and} \quad [\,\mb z'_{\bfthres}\;|\;\dmpar], 
[\mb z'_i\,|\,\dmpar] \;\;(1\le i\le\nblock)
$$

\noindent composing the     augmented likelihood
\eqref{eq:augmentedLKL},    according to 
 equations \eqref{eq:jointCompletedDensity} and
\eqref{densityZu}, and let the factors 
 $\varphi_{\bfthres}, \;\varphi_{i},\;i\le \nblock$ of the 
weight function $\varphi$ % completing the
% definition of the augmented posterior
in \eqref{eq:InveriantMeas} be defined by equation~\eqref{defEf}.

Then, the augmented posterior $[\,\mb z,\fullpar\,|\,\mb O\,]\aug
\propto  [\,\fullpar\,][\mb z,\mb O\,|\,\fullpar\,]\,\varphi(\mb z)\  $ is consistent, in the sense that 
 condition~\eqref{eq:consistencyLikelihood}, whence
\eqref{eq:consistencyAugmented},  is satisfied.

\end{proposition}

% \begin{equation*}
%   [\,\mb z_{t_i} , \mb 0 \, | \,\theta] = 
% \end{equation*}
\paragraph{\textbf{\textit{Proof}}}\ \\ 
It is enough to   show that, on the one hand, 

\begin{equation}
  \label{eq:consistAbove}
\int [\mb z_{\text{above}},\mb O\;|\;\fullpar\,] \ud \mb
z_{\text{above}} =  
\prod_{i=1}^{n_{\bthres}} \Bigg\{
   \int_{\mb{[\tilde L}_{t_i},\mb{\tilde R}_{t_i}\mb]}
  \frac{\ud \expmeas_{\dmpar}}{\ud \mb x}
\ud\genemeas_i(\mb x) \prod_{ j: y_{j,t_i} > \thres_j} J_j^{\margpar}(y_{j,t_i}) \,,
\Bigg\}  
\end{equation}

 and, on the other hand, 

\noindent
\begin{equation}\label{integratedAugmentedExpfun}
\begin{cases}
 \PE\left[
\varphi_{{\bfthres}}(\mb Z'_{\bfthres})\right]   = 
 \exp\left(-\ndef\,
\expmeas_{\dmpar}\left(A_{\bfthres} \right) \right)\,,
\\
\PE\left[
\varphi_{{i}}(\mb Z'_{i})\right]   = 
 \exp\left(-n'_i\,\expmeas_{\dmpar}(A'_i)
 \right)&(i\le \nblock)\,.
\end{cases}
\end{equation} 

\noindent where the above expectations are taken  with respect to
$[\mb Z'_{\bfthres}\,|\,\dmpar\,]$ and $[\mb Z'_{i}\,|\,\dmpar\,]$.

Establishing \eqref{eq:consistAbove} is immediate: from
definition~\eqref{eq:jointCompletedDensity} of the $[\mb z_{t_i}, \mb
O|\theta] $'s composing $[\mb z_{\text{above}}, \mb O|\theta]$,

\begin{equation*}
\int_{\mb{[\tilde L}_{t_i},\mb{\tilde R}_{t_i}\mb]}
[\mb z_{t_i  }, \mb O_{t_i}|\theta] \ud \mb z_{t_i} 
= 
   \int_{\mb{[\tilde L}_{t_i},\mb{\tilde R}_{t_i}\mb]}
  \frac{\ud \expmeas_{\dmpar}}{\ud \mb x}
\ud\genemeas_i(\mb x) %\times\dotsb
\prod_{ j: y_{j,t_i} > \thres_j} J_j^{\margpar}(y_{j,t_i}) \,,
\end{equation*}

\noindent
which yields~\eqref{eq:consistAbove} by taking the product over
indices  $1\le i\le n_{\bthres}$. 

% Thus for  the consistency condition \eqref{eq:consistencyLikelihood} to
% hold, it is   that 
It remains to show \eqref{integratedAugmentedExpfun}. 
% Since the  
%    $\mb Z'_{\bfthres}, \mb Z'_{i},\,(i\le \nblock)$ are 
% independent Poisson processes, with respective densities  $[\mb z'_{\bfthres}|\dmpar]$, 
% $[\mb z'_{i}|\dmpar],$ 
% and  that  $\varphi_{\bfthres} (\mb z'_{\bfthres}), \varphi_{i}
% (\mb z'_{i})$, satisfy 
% % To enforce \eqref{integratedAugmentedExpfun}, 
% In particular,  $Z'_{\bfthres}$ and the
% $Z'_{i}$'s may be defined
%   as independent  Poisson processes and the Laplace transform
% % of  Poisson processes 
% plays a central role in the choice of 
% $\varphi$.  
To wit, $\varphi_{\bfthres}$ is a smoothed version of the indicator
function $\mathds{1}_{\{\mb N'_{\bfthres}(A_{\bfthres,\ndef}) = 0\}}$,
which expectancy is $\PP( \mb N'_{\bfthres}(A_{\bfthres,\ndef}) = 0) =
e^{-\ndef\expmeas_\dmpar(A_{\bfthres})}$, as soon as $\mb N'_{\bfthres}$ is a Poisson
process with intensity measure $\expmeas_\dmpar$.
For $f$  a bounded, continuous
  function defined on a nice space $E$ and $N$ a point process on $E$,
   denote

\noindent 
$$N(f) = \int_{E} f\ud N  = \sum_{i=1}^{N(E)} f(s_i).$$

\noindent
 Then, if $N$ is a Poisson process $\PRM(\expmeas)$
  on $E$, the Laplace transform $\mathit{Lap}_{N}(f) \hat{=}
  \mathbb{E}(e^{-N(f)}) $ is \citep[Chap. 3]{resnick1987extreme}

\noindent
$$
\mathit{Lap}_{N}(f) = \exp\left(-\int_{E} (1-e^{-f(s)})
  \ud\expmeas(s)\right)\, .
$$

Consider the region   $E = E_{\bfthres,\ndef}$  as in \eqref{defEf}  % be any region of $\mb E$ containing
and take 

\noindent
$$
f_{\bfthres}(\mb x) =
-\log(1-1/\tau)\mathbf{1}_{A_{\bfthres,\ndef}}(\mb x)\,,\quad
\text{ so that }
\quad 
1-e^{-f_{\bfthres}} = \frac{1}{\tau}\mathds{1}_{A_{\bfthres,\ndef}}\,.
$$
\noindent
With these notations, 

\noindent
$$\varphi_{\bfthres}(\mb Z'_{\bfthres}) \;\hat= \; 
\left(1-\frac{1}{\tau}\right)^{\mb Z'_{\bfthres}(A_{\bfthres,\ndef})} \; = \; 
\exp\left(-\mb Z'_{\bfthres}(f_{\bfthres})\right) , $$

\noindent
 whence

\noindent
$$
\begin{aligned}
\PE\left(\varphi_{\bfthres}(\mb Z'_{\bfthres} )\right)  &  = 
\mathit{Lap}_{\mb Z'_{\bfthres}}(f_{\bfthres}) \\
& = \exp\left(-\int_{E_{\bfthres,\ndef}} (1 - e^{-f_{\bfthres}}) \ud\lambda'\right) \\
& = \exp\left(-\int_{E_{\bfthres,\ndef}}
\frac{1}{\tau}\mathds{1}_{A_{\bfthres,\ndef} }\ud (\tau
\lambda_\dmpar) \right)\\
& =\exp\left(- \lambda_\dmpar(A_{\bfthres,\ndef}) \right)\,.
\end{aligned}
$$

\noindent This shows the first equality in 
\eqref{integratedAugmentedExpfun}. The second one is  derived
with  a similar argument. 

\ \, \hfill $\square$

% It is enough to construct  a function $f_{\bfthres}$ on
%  $E_{\bfthres,\ndef}$  and an intensity  measure 
% $\expmeas'$ such that 
% \begin{equation}\label{goodf}
% \int_{E_{\bfthres,\ndef}} (1-e^{-f_{\bfthres}(s)})
%   \ud\expmeas'(s) = \expmeas_{\dmpar}(A_{\bfthres,\ndef})\,.
% \end{equation}
% %Indeed, suppose \eqref{goodf} holds, for some $f$. 
% Indeed, letting

% \begin{equation*}
% %  \label{eq:ppAugment}
%   \begin{cases}
% \mb Z'_{\bfthres}\sim PRM(\expmeas') \text{ on } E_{\bfthres,\ndef}\,,\\
% \varphi_{\bfthres}(\mb Z'_{\bfthres}) =  \exp(-\mb Z'_{\bfthres}(f_{\bfthres}))\,,
%  \end{cases}
% \end{equation*}
% one has  $\mathbb{E}\big[\varphi(\mb Z'_{\bfthres})|\dmpar\big] =
% \mathit{Lap}_{\mb Z'_{\bfthres}}(f_{\bfthres}) = 
% %% \mathbb{E}(\exp(-N(f)|\dmpar) =
% \exp(-\expmeas_{\dmpar}(A_{\bfthres,\ndef}))$, as required by
% condition~\eqref{integratedAugmentedExpfun}.
% In order to satisfy  \eqref{goodf}, let us  fix a
% multiplicative constant $\tau >1$ and define 
% \begin{equation*}
% %\label{defEf}
% \begin{cases}
% % E_{\bfthres,\ndef}= \{\mb x \in (\mathbb{R}^+)^d : \|\mb x\|_1 > 
% % \min_j(\frac{\fthres_j}{\ndef} ) \}\;, \\
% \expmeas'(\point) = \tau \,\expmeas_{\dmpar}(\point)\;,  \\
% f_{\bfthres}(\mb x) = -\log(1-1/\tau)\mathbf{1}_{A_{\bfthres,\ndef}}(\mb x)\,.
% \end{cases}
% \end{equation*}
% % where $\mathbf{1}_E(\point)$ denotes the characteristic function of
% % the set $E$. 
% %Thus $A_{\bfthres,\ndef} \subset  E_{\bfthres,\ndef}$, and 
% Then,  %$ \eqref{defEf}~\Rightarrow~\eqref{goodf}$. 
% \eqref{goodf} holds.

The points of $\mb Z'_{\bfthres}$ can easily be simulated  \citep[see][
Chap.3]{resnick1987extreme}:  the number  of points $N'_{\bfthres}$ in
$E_{\bfthres, \ndef}$ is a
Poisson random variable with mean equal to 

$$
\expmeas'(E_{\bfthres,\ndef})~=~\frac{\tau~\dimens~}{\min_j(\fthres_j~/\ndef)}\,,
$$

\noindent
and each point  has density  in polar coordinates   equal to 
 $\frac{1}{\lambda'(E_{\bfthres,\ndef})}\frac{\tau d}{r^2} h_{\dmpar}(\Ang)$.

% To sum up, the  desired functional $\varphi_{\bfthres}$ is  

% $$\varphi_{\bfthres}(\mb Z'_{\bfthres}) = e^{\log(1-1/\tau)
%   \mb Z'_{\bfthres}(A_{\bfthres,\ndef}) }  = 
% (1-1/\tau)^{   \mb Z'_{\bfthres}(A_{\bfthres,\ndef})}\,,$$
% as it appears  in equation~\eqref{defEf}.
% where $Z'_{\bfthres}(A_{\bfthres,\ndef})$ is the random number of  points forming
% $Z'_{\bfthres}$ which fall
% in $A_{\bfthres,\ndef}$.
% For  $\tau$ close to one, $\varphi'_{\bfthres}(\mb Z'_{\bfthres})$ is close to
% $\mathbf{1}_{\mb Z'_{\bfthres}(A_{\bfthres,\ndef})=0}$.

\begin{rem}
  One may be tempted to define $\mb Z'_{\bfthres}$ 
as a  
Poisson process with intensity $\lambda_{\dmpar}$ on some 
$E\supset A_{\bfthres,\ndef}$, and  $\varphi(\mb Z_{\bfthres})$ 
as the indicator  
$\mathds{ 1}_{\mb z'_{\bfthres}(A_{\bfthres,\ndef}) = 0}$, 
with a similar definition for the 
$\varphi_i'$'s and the $\mb Z'_i $'s. As pointed out in the proof of
Proposition~\ref{prop:consistency}, 
one would have 
$\mathbb{E}\left[
\mathbf{1}_{\mb Z'_u(A_{\bfthres,\ndef}) =0}  \right]= 
\exp\left(-\ndef\,
\expmeas_{\dmpar}\left(A_{\bfthres} \right) \right)$, as required. 
However, even if this construction is valid in theory, it leads to a
very large rate of rejection in the Metropolis algorithm: $\varphi(\mb
Z'_{\bfthres})$
has too much variability around its mean value and the proposal
is systematically  rejected each time a point in the augmentation
process  hits  the
failure region. 

% Here is detailed the  alternative  construction of
% $ \mb Z'_{\bfthres}$ and $\varphi_{\bfthres}$ which is used in this
% paper (the construction for the $\mb Z'_i, i\le \nblock$ and the
% $\varphi_i$'s is similar).
\end{rem}

\subsection{Expression of the augmented posterior }
\label{express_augmentedPos}
Recall from Section~\ref{sec:DataAugmentPoissonModel}, equation~\eqref{eq:InveriantMeas}, that  
the augmented posterior density to be sampled by the \MCMC\ algorithm is 

\noindent
$$
[\, \mathbf{z},
\fullpar\, |\, \mb O\,]\aug \propto  
[\,\fullpar\, ]
[\, \mathbf{z}, \mb O \,|\,
\fullpar\, ]\varphi(\mb z)\,.
$$

\noindent Combining equations~\eqref{eq:jointCompletedDensity}
and  \eqref{defEf}, and integrating out missing components as in
Appendix~\ref{sec:integrMissing}, the developed expression is 
\begin{multline}\label{lkl_total}
[\,\mathbf{z},
\,\fullpar \, |\, \mb O\,]\aug \;\propto\; [\,\theta\,]\;
\left([\mb z'_{\bfthres}|\dmpar]\prod_{i=1}^{\nblock} [\mb z'_i|\dmpar]
\right)\point
(1-1/\tau)^{\mb z'_{\bfthres}(A_{\bfthres,\ndef}) + 
    \sum_{i=1}^{\nblock} \mb z'_i(A'_{i,n'_{i}})}\,
%
% \frac{1}{
% Z'_{\bfthres}(E_{\bfthres,\ndef})!
% \prod_{i=1}^{\nblock}Z'_{i}(E_{i,n'_i})!
% }
% %%
% e^{\frac{-\ndef\,\tau\,\dimens  }{\min_{j\le d}(\fthres_{j})} -
%  \sum_{i=1}^{\nblock}
% \frac{n'_i\,\tau\,\dimens  }{\min_{j\le d}(\mb{\tilde{R}}_{j,t'_i})}}
%  \times\\
% \dotsb\prod_{s= 1}^{Z'_{\bfthres}(A_{\bfthres,\ndef})}
% \frac{\tau\,\dimens}{(R'_{\bfthres,s})^2} h_{\dmpar}(\mb W'_{\bfthres,s}%)\,.
% %%\frac{\ud\expmeas_{\dmpar}}{\ud \mb x } ({\mb X}'_{\bfthres,s})
%  \times
% \prod_{i=1}^{\nblock}\Big\{\prod_{s= 1}^{Z'_i(A'_{i,n'_{i}})}
% \frac{\tau\,\dimens}{(R'_{i,s})^2} h_{\dmpar}(\mb W'_{i,s})\,.
% %\frac{\ud\expmeas_{\dmpar}}{\ud \mb x } ({\mb X}'_{i,s})
% \Big\}
  \dotsb\\
\dotsb \prod_{i=1}^{n_{\bthres}} \Big\{
\frac{\partial^{r(i)} \expmeas_{\dmpar}}{\partial x_{j_1(i)}\dotsb
  x_{j_{r(i)}(i)}} (\mb{\bar{x}}_{t_i})
%  \frac{\ud\expmeas_{\dmpar}}{\ud \mb x } (\tilde{\mb x}_i)
\prod_{ j: Y_{j,t} > \thres_j} J_j^{\margpar}(y_{j,t})
 \Big\}\,,
\end{multline}

\noindent where $[\mb z'_{\bfthres}|\dmpar]$ and the $[\mb
z'_{i}|\dmpar]$'s are given by equation~\eqref{densityZu}. 
\section{\MCMC\ algorithm}\label{sec:MCMC algo}

The MCMC algorithm generates a sample $(\theta_\iota, \mb
Z_\iota)_{\iota = 1,\dotsc,N}$ which distribution converges to the
invariant distribution of the chain, which is the augmented posterior
distribution $[\mb Z,\theta | \mb O]\aug$, as defined in
Section~\ref{sec:data_aug} and Appendix~\ref{express_augmentedPos}.  The quantity of interest here is
the %unconstrained parameter is denoted $\dmupar$, and the
joint parameter $\fullpar$, which is the concatenation of the marginal
parameters and the %unconstrained
dependence parameter: $\fullpar = (\margpar, \dmpar)$. We recall that
 MCMC algorithms aiming at sampling a quantity $\Delta\in E$
 according to a density $\pi(\point)$
proceed typically as follows
\begin{itemize}
\item Start with any value $\delta(0)\in E$
\item for $\iota\in\{1,\dotsc,N\}$
  \begin{enumerate}
  \item Generate $\delta^*$ according to a \emph{proposal distribution}
    with density $q(\delta(\iota),\point)$
  \item Compute the \emph{acceptance ratio} 

\noindent
$$
\alpha =
\frac{\pi(\delta^*)\;q(\delta^*,\delta(\iota))}{
\pi(\delta(\iota))\;q(\delta(\iota),\delta^*)}\;,
$$

\noindent and generate $U$, a uniform random variable on $[0,1]$.
\item If $U>\alpha$, `reject' the proposal and  set 

\noindent
$$
\delta(\iota+1) = \delta(\iota)\,. 
$$

\noindent 
Otherwise (\ie with probability $\alpha$), set 

\noindent
$$
\delta(\iota+1) = \delta^*
$$
\item Set $\iota = \iota+1$, go to (1).

  \end{enumerate}
\item Return $\Big(\delta(\iota_{\min}), \dotsc, \delta(N)\Big)$,
\end{itemize}
where $\iota_{\min}$ is the length of the \emph{burn-in period},
 after which  the chain is deemed to have reached a stationary
 behavior. 
In our case, % For each particular move, the acceptance ratio is set  in such a way,
% that 
the
unnormalized posterior measure  
$[\mb Z,\theta|\mb O]\aug$ (see equation \eqref{lkl_total}) on the augmented
parameter space plays the role of the objective density $\pi$ above.

In a \emph{Metropolis-within-Gibbs} MCMC, several proposal kernels
$q_1,\dotsc, q_T$ are
defined, each of them corresponding to  a \emph{type of move}, which is randomly chosen
among $\{1,\dotsc, T\}$ at each iteration $\iota$ and allows to modify
some subset of  components in $\delta$ alone.  The algorithm developed in this paper builds on the MCMC
algorithm proposed by \cite{sabourinNaveau2012}, in which several \emph{types of
move} modifying the dependence structure (Dirichlet mixture parameter
$\dmpar$) have been defined. Those are kept as  they are in the present
work, the novel part of which concerns the definition of
\emph{marginal moves} (modifying the marginal parameter $\margpar$)
and \emph{augmentation moves} (modifying the augmentation data $\mb Z$). 
% proposals generated by the algoritm  are denoted with a $\ ^*$ and the
% current state of the chain after 
% $\iter$ iterations is a function of  $\iter$.  The
% proposal kernels are  $Q(\theta(\iter), \point)$ and their density % (in the case of
% % continuous proposals)
% are  $q(\theta(\iter),\point)$. The acceptance ratios are denoted
% $\alpha$.
Additional notations distinguishing between the quantities appearing
in the general MCMC algorithm above, according to the type of move,
are omitted in the remainder of this section. 

\subsection{Starting values} In a preliminary step, likelihood optimization is performed in the
independent model (the likelihood for one multivariate observation is
the product of $\dimens$ Pareto densities). This provides starting
values for the marginal parameters as well as a Hessian matrix $\hess$,
that may be used as the inverse of a reference covariance matrix when
updating the marginal parameters.

\subsection{Marginal moves} 
The marginal parameter $\margpar$  is updated as a block: 
The proposal is normal, with mean at  $\margpar(\iter)$ and co-variance
matrix  $\Sigma = \delta\, \hess^{-1}$, where $\delta$ is a scaling factor
fixed by the user, that  may  typically be set around $0.5$ and
$\hess$ is the  Hessian matrix computed in the preliminary step. 
 Since the proposal density is symmetric, and since   this move does not modify the dependence structure,
 neither the terms involving the proposal density, nor the  point
 processes $\mb Z'$,   appear in acceptance
ratio. The augmented variables $\mb Z_{\text{above}}$
% components $\bar{X}_{j,t_i} : \tilde \datatype_{j,t_i} \in \{2,3 \}$ 
are left unchanged. If any augmented component $\mb Z_{j,t_i}$ is
outside of the candidate censoring interval $\mb [\mb{\tilde L}_{t_i}^*
\mb{, \tilde R}_{t_i}^* \mb ]$ (on the new Fréchet scale) resulting from the modification of the
marginal parameters, the move is rejected, since  in such a case,  the candidate has
augmented likelihood $[\mb z, \mb O\,|\, \theta^* \,]= 0$, \ie $\alpha=0$. 

Otherwise,  the uncensored Fréchet-transformed variables (such that $\tilde{\datatype}_{j,t_i} = 1$) are updated
to 
$ \mb{{ X}}_{t_i}^* = 
\mathcal{T}_{j,t_i}^{\margpar^*}(Y_{j,t_i})
$. The acceptance ratio is 

\begin{multline*}
\alpha = \frac{ [\margpar^*]}{[\margpar(\iter)]}
\prod_{i=1}^{n_{\bthres}} \Bigg\{
  \frac{
    \partial^{r(i)} \expmeas_{\dmpar}  
  }{
    \partial x_{j_1(i)}\dotsb
   \partial x_{j_{r(i)}(i)
    }
  }
  (\mb{\bar{ X }}^{*}_{t_i})
%%}{
\left[
  \frac{
    \partial^{r(i)} \expmeas_{\dmpar}
  }{
    \partial x_{j_1(i)}\dotsb
    \partial x_{j_{r(i)}(i)}
  }
  (\mb{\bar  X}_{t_i}(\iter))
%}
\right]^{-1}
% \Big. \\
% \Big.
\\
\prod_{ j: Y_{j,t_i} > \thres_j} 
\frac{J_j^{\margpar^*}(Y_{j,t_i})}
{J_j^{\margpar(\iter)}(Y_{j,t_i}) }
\Bigg\},
\end{multline*}

\noindent
where $j_{1}, \dotsc,j_{r(i)}$ are the non-missing components in the
censored observation $\mb C_{t_i}^\margpar$ and $\bar X_j = X_j$ for
uncensored components, $\bar X_{j,t_i} = Z_{j,t_i}$ otherwise (see Appendix~\ref{sec:integrMissing}). 

\subsection{Augmentation moves for $\mb Z_{\text{above}}$}
The augmented components $\{Z_{j,t_i}\} = \{\bar{\mb X}_{j,t_i} :
\tilde\kappa_{j}\in\{2,3\} \}$
%\mathscr{D}_c(i) \}$
(\emph{c.f.} Section~\ref{sec:DataAugmentPoissonModel}) are  re-sampled, one
coordinate at a time,  from their exact 
conditional distribution given the other coordinates, as derived in
Appendix~ \ref{sec:dataAugmentDetails}. %  More details
% about the sampling procedure are gathered in appendix~\ref{ap: augmentedDistr}.
% Again, the proposal parameters are directly sampled from their full
% conditional distribution, and no
Since no other component of $(\mb Z,\theta)$ is modified, the proposal
density equals the objective density, and the acceptance ratio
is thus set to $\alpha=1$.

\subsection{Augmentation moves for $\mb Z'$}
During this move, proposals 

\noindent
$$
\mb Z'^* = \left\{ \mb{Z}'^*_{\bfthres}, \mb{Z}'^*_i, \;
i\le \nblock \right\} \,,
$$  

\noindent
for the   augmentation  Poisson processes  introduced  in the end of Section~\ref{subsec:adaptCensor}, 
are sampled under their exact  distribution,  

\noindent
$$
q\Big( \mb Z'(\iter),\mb Z'^*\Big)  =  [\mb Z'^*|\dmpar(\iter)]\,.
$$

 The latter is  determined by
their intensity measure $\lambda' = \tau \expmeas_{\dmpar}$:  the
multiplicative constant $\tau$
 and the sampling procedure have been  described  in
Appendix~\ref{sec:augmentExp}. The acceptance ratio is thus 

$$
\alpha = \frac{[\,\mb Z'^*,\theta(\iota)\,|\,\mb O\,]\aug  \;\;
[\,\mb Z'(\iota) \,|\, \dmpar(\iter)\,] }{
[\,\mb Z'^*,\theta(\iota)\,|\, \mb O\,]\aug\;\; 
[\,\mb Z'^* \,|\, \dmpar(\iter)\,]
}
$$

% Since their distribution depends  on
% $\dmpar$ only, 
% the acceptance ratio is

\noindent all the terms cancel out except the ratio
$\varphi(\mb Z'^*)/\varphi(\mb Z'(\iter))$, so that 

$$
\alpha = (1-1/\tau)^{\left[(\mb Z'_{\bfthres})^*(A_{\bfthres,\ndef}) -
\mb Z'_{\bfthres}(\iter)(A_{\bfthres,\ndef}) 
 \right] + 
\sum_{i\le \nblock} 
\left[(\mb Z'_{i})^*(A'_{i,n'_i}) -
\mb Z'_{i}(\iter)(A'_{i,n'_i})  
 \right] }\,.
$$
\subsection{Dependence moves}
These types of moves allow to update $\dmpar(\iter)$.% $\dmupar(\iter)$ (and consequently also
% the original dependence parameter $\dmpar(\iter)$.
% As in \cite{sabourinNaveau2012}, six dependence moves are defined: 
% $\mu$-moves, $\nu$-moves, $\ecc$-moves, split-moves, combine-moves and
% shuffle moves.
The only  difference between the present algorithm and what is
described is  \cite{sabourinNaveau2012} is that 
not enough exact  angular data are available to construct proposals for
moving or 
splitting a  Dirichlet mixture components $\Mu_{ m}$ .
% during the $\mu$-moves and the split-moves as Dirichlet kernels for
% the proposal distribution of some candidate $\Mu_{  m}^*$ (where $m= k(\iter)$ for a split move).
Indeed, most of the  observations have at least
one coordinate missing or censored, so that no `angle' is
available. Consequently, the latter proposal %  for moving or splitting one 
% Dirichlet location parameter
is a  simple Dirichlet 
distribution with mode at $\Mu_{ m}(\iter)$, with re-centering
parameter $0< \epsilon < 0.5$,

\noindent
$$
q(\Mu_{m}(\iter),\point) = \diri_{\frac{\dimens}{\epsilon},
  \gamma^{*} }(\point)\, ,
$$

\noindent
with $\gamma^{*} = (1-\epsilon)\,\Mu_{m} + 
\epsilon\,(\frac{1}{\dimens},\dotsc,\frac{1}{\dimens})$. 

Each dependence move (except for   a \emph{shuffling move} which only affects
the representation of the angular distribution, see \cite{sabourinNaveau2012})  is systematically 
followed by an augmentation move updating the Poisson processes $\mb
Z'$, %  move, 
which improves the chain's
 mixing properties. This  also  avoids  the computation of  the `costly'
term involving the density $[\mb z' |\dmpar]$ (see
equation~\eqref{densityZu}). % of the points forming  $\mb Z'$ in the likelihood
% ratio:
 Indeed, the acceptance
ratio for the two consecutive moves (dependence move  followed by a 
augmentation move) is 

\begin{multline*}
\alpha = \frac{[\dmpar^*]}{[\dmpar(\iter)]}
\frac{q(\dmpar^*,\dmpar(\iter))}{q(\dmpar(\iter),\dmpar^*)}\,
 \times\dotsb\\
 \dotsb(1-1/\tau)^{\left\{
\left[\mb Z'^*_{\bfthres}(A_{\bfthres,\ndef}) -
\mb Z'_{\bfthres}(\iter)(A_{\bfthres,\ndef}) 
 \right] + 
 \sum_{i\le \nblock} 
 \left[\mb Z'^*_{i}(A'_{i,n'_i}) -
 \mb Z'_{i}(\iter)(A'_{i,n'_i})  
  \right]
\right\}}\,
\times\dotsb\\
%(1-1/\tau)^{Z'^*(A_{\bfthres,n}) - Z'(\iter)(A_{\bfthres,n})}\,\\
\dotsb\prod_{i=1}^{n_{\bthres}} \Bigg\{
  \frac{
    \partial^{r(i)} \expmeas_{\dmpar^*}  
  }{
    \partial x_{j_1(i)}\dotsb
   \partial x_{j_{r(i)}(i)
    }
  }
  ({\mb{\bar{ X}}}_{t_i}(\iter))
\left[
  \frac{
    \partial^{r(i)} \expmeas_{\dmpar(\iter)}
  }{
    \partial x_{j_1(i)}\dotsb
    \partial x_{j_{r(i)}(i)}
  }
  ({\mb{\bar{X}}}_{t_i}(\iter))
\right]^{-1}
\Bigg\}.
\end{multline*}

\subsection{MCMC settings and convergence diagnostics  in  the simulation study}
\label{sec:MCMCsimstudy}

For the simulation study, the   prior on the Dirichlet mixture distributions is
specified in a similar way as in
 \cite{sabourinNaveau2012}. 
 The number $k$ of mixture
components has truncated geometric distribution, 
$[k] \propto \left(1-\frac{1}{\lambda}\right)^{k-1}\frac{1}{\lambda}
% = \mathrm{Poiss}(k | \lambda)
 \mathds{1}_{[1, k_{\max}]}(k)
$
with upper bound
$k_{\max} = 10$ and mean parameter $\lambda=4$.
Also, for the sake of simplicity, all the marginal parameters  are assumed
to be \emph{a priori} independent, with normal distributions (after
log-transformation of the scales). The shape parameter has standard
normal distribution and the logarithms of the scales have mean and
standard deviation both equal to $5$. 

As for the  augmentation Poisson process data, the multiplicative constant $\tau$
involved in the Poisson intensity  is set to $50$. It
appeared that smaller values of $\tau$ (close to $1$) considerably affected the
mixing properties of the chains. 

 Convergence of the dependence parameters
$\dmpar(\iter)$  can be 
monitored  using  functionals 
 based on integration of the simulated densities against
Dirichlet test functions \citep[see][for
details]{sabourinNaveau2012}. To detect possible mixing defects, six
chains of $10^6$ iterations each are   run in parallel.  
Standard convergence diagnostic tests are implemented in \textsc{R} 
\citep{heidelberger1983simulation,gelman1992inference}, respectively
testing for non-stationarity and poor mixing. For example, the stationarity test 
 detects  three  
non-stationary chains out of six for the  simulated
data set exemplified in Section~\ref{sec:OneSimDataset}.
 The mixing properties of the three  retained
ones, as measured by a variance ratio inter/intra chains, 
are satisfactory enough:  all the potential scale reduction factors
\citep{gelman1992inference} are below $1.1$. The same is true of the
marginal parameter component of the chains, 
% As for the marginal parameter components 
$(\margpar(\iter) )_\iter$. 
\bibliographystyle{apalike}
 \bibliography{censoredExtremes}
%\bibliography{Censoredextremes141128.bbl}
\vfill

\end{document}